\documentclass[pra,aps,twocolumn,nopacs,superscriptaddress,nofootinbib]{revtex4}
\usepackage{graphicx}
\usepackage{dcolumn}
\usepackage{bm}
\usepackage{amsmath}
\usepackage{epsfig}
\usepackage{braket}
\usepackage{slashbox}
\usepackage{bbm}
\usepackage{color}

\begin{document}

\def\ii{{i}}  \def\ee{{\rm e}}  \def\vb{{\bf v}}
\def\Rb{{\bf R}}  \def\pb{{\bf p}}  \def\Qb{{\bf Q}}  \def\ub{{\bf u}}
\def\jb{{\bf j}}  \def\kb{{\bf k}}  \def\Eb{{\bf E}}  \def\rb{{\bf r}}
\def\xx{\hat{\bf x}}  \def\yy{\hat{\bf y}}  \def\zz{\hat{\bf z}}
\def\th{\vec{\bf{\theta}}}
\def\fE{\vec{\mathcal{E}}}  \def\Mb{{\bf M}}  \def\EF{{E_{\rm F}}}
\def\vF{{v_{\rm F}}}   \def\EFT{{E_{\rm F}^T}}  \def\kB{{k_{\rm B}}}

\def\Brack#1{\left[ #1 \right]}
\def\bra#1{\mathinner{\langle{#1}|}}
\def\ket#1{\mathinner{|{#1}\rangle}}
\def\braket#1{\mathinner{\langle{#1}\rangle}}
\def\Bra#1{\left<1>}
{\catcode`\|=\active\gdef\Braket#1{\left<\mathcode`\|"8000\let|\bravert {#1}\right>}}
\def\bravert{\egroup\,\vrule\,\bgroup}

\title{Ultrafast radiative heat transfer}
\author{Renwen~Yu}
\affiliation{ICFO-Institut de Ciencies Fotoniques, The Barcelona Institute of Science and Technology, 08860 Castelldefels (Barcelona), Spain}
\author{Alejandro~Manjavacas}
\affiliation{Department of Physics and Astronomy, University of New Mexico, 1919 Lomas Blvd. NE, Albuquerque NM 87131-0001, United States}
\author{F.~Javier~Garc\'{\i}a~de~Abajo}
\email{javier.garciadeabajo@icfo.es}
\homepage{http://www.nanophotonics.es}
\affiliation{ICFO-Institut de Ciencies Fotoniques, The Barcelona Institute of Science and Technology, 08860 Castelldefels (Barcelona), Spain}
\affiliation{ICREA-Instituci\'o Catalana de Recerca i Estudis Avan\c{c}ats, Passeig Llu\'{\i}s Companys 23, 08010 Barcelona, Spain}

\begin{abstract}
{\bf Light absorption in conducting materials produces heating of their conduction electrons, followed by relaxation into phonons within picoseconds, and subsequent diffusion into the surrounding media over longer timescales. This conventional picture of optical heating is supplemented by radiative cooling, which typically takes place at an even lower pace, only becoming relevant for structures held in vacuum or under extreme conditions of thermal isolation. Here we reveal an ultrafast radiative cooling regime between neighboring plasmon-supporting graphene nanostructures in which noncontact heat transfer becomes a dominant channel. We predict that $>$50\% of the electronic heat energy deposited on a graphene disk can be transferred to a neighboring nanoisland within a femtosecond timescale. This phenomenon is facilitated by the combination of low electronic heat capacity and large plasmonic field concentration displayed by doped graphene. Similar effects should take place in other van der Waals materials, thus opening an unexplored avenue toward efficient heat management in ultrathin nanostructures.}
\end{abstract}
\maketitle

\section{Introduction}

Optical, electrical, and mechanical dissipation in nanoscale devices produces heat accumulation that can result in structural damage and poor performance. Understandably, heat management constitutes an important aspect when designing thermoelectric \cite{VSC01}, optoelectronic \cite{TZK07}, electromechanical \cite{SGZ14}, and photovoltaic \cite{WZM03} elements, as well as recently proposed thermal analogs of electronic devices \cite{BB14,KBB14}. However, the relatively slow thermal conduction in most materials \cite{L05} imposes a serious limitation. Finding new means of cooling nanostructures is therefore critical. An interesting possibility is provided by coupling to radiative degrees of freedom. Indeed, the absorption and emission of radiation by a material structure contributes to reach thermal equilibrium with other surrounding structures and the electromagnetic environment. This is the dominant cooling channel for thermally isolated structures \cite{GOP12}, in which energy is released through the emission of photons with wavelengths $\sim\lambda_T=2\pi\hbar c/\kB T$ (i.e., the thermal wavelength at temperature $T$). When the structures are separated by vacuum gaps of large size compared with $\lambda_T$, the Planck and Kirchhoff laws determine the exchanged power \cite{R1965}. In contrast, for neighboring objects separated by a small distance compared with $\lambda_T$, radiative heat transfer is dominated by additional channels mediated by evanescent waves \cite{H1969,DBT1970,PV1971}. These can produce rates exceeding the black-body limit by several orders of magnitude, enhanced by near-field coupling of resonances supported by the nanostructures, thus emerging as a potentially relevant transfer mechanism in solid state devices.

Following pioneering observations of near-field radiative energy transfer between two conducting plates \cite{H1969,DBT1970}, a theoretical explanation was offered \cite{PV1971} based on the effect of thermal fluctuations in the electrical current of the involved surfaces. Further experimental \cite{KMP05,DFC06,NSH09,RSJ09,SNC09,OQW11,JR12,KSF15,
STF16,SZF16} and theoretical \cite{LM94,CG99,P99,SJC00,MJC01,
VP01,NC03,VP04,DVJ05,VP07,CLV08,CLV08_1,M08,NC08,PRL08,
BZF09,PLR09,BJ10,DK10,PM10,BBJ11,KEK11,MA11,OF11,RIB11,
GCM12,GLR12,paper198,MTB13,BB14,LZZ14,MJR15} studies have corroborated this interpretation of radiative heat transfer between structures of varied morphologies. This subject has generated fundamental insights that include important corrections due to nonlocal \cite{VP04}, phonon \cite{MJC01,PM10}, and photonic band \cite{RIB11} effects, as well as magnetic polarization \cite{CLV08_1}. Additionally, retardation, radiation emission, and crossed electric-magnetic terms in the optical response have been shown to severely modify the transfer power \cite{paper198}. However, the so far observed and predicted transfer rates are slow compared with dissipative transport through the surrounding media, in which heat can cause undesired effects. This situation persists even when the interaction between neighboring structures is enhanced due to strong resonant excitations, such as plasmons in noble metals.

In this context, graphene plasmons can be advantageous because their frequencies lie in the mid-infrared, which is the spectral region for thermal interactions under attainable temperatures. Indeed, plasmon energies in graphene nanostructures scale as $\sim\sqrt{\EF/D}$ with the Fermi energy $\EF$ and the characteristic size $D$ (e.g., the diameter for a disk). Doping levels as high as $\EF\sim1\,$eV have been reported through electrostatic gating \cite{CPB11}, and even higher values through chemical doping \cite{LLZ11,KWB12}, manifesting themselves in the opening of a $2\EF$ gap for vertical optical transitions\cite{LHJ08,CPB11}. However, plasmons are only well defined at energies below $\sim\EF$ due to the narrowing of the gap as their momentum increases \cite{paper235}. For reference, a 20\,nm disk supported on silica and doped to $\EF=1\,$eV exhibits a dipolar plasmon at $\approx0.4\,$eV \cite{paper235}. This explains why experiments have only explored mid-infrared plasmons, as higher energies require smaller structures, whose fabrication can be challenging.

An additional advantage of graphene lies in its large electrical tunability, which enables an active control of these phenomena. In a related context, electrical modulation of thermal emission of radiation has been accomplished in gated nanostructured graphene \cite{BSJ15}, while an optical-to-thermal converter has been proposed to be capable of efficiently transforming an optical pump into light emitted at longer mid-infrared wavelengths \cite{paper245}. Electrical control of radiative heat transfer between graphene-coated surfaces or between extended graphene and other materials has been also proposed \cite{PV11,IJJ12,VTB12,SVC12,LZ14}.

The competing mechanism (relaxation into phonons) was initially thought to be rather slow in graphene \cite{BM09} (nanosecond scale), a prediction that was subsequently corrected to much shorter timescales (picoseconds) due coupling of hot charge carriers to optical phonons \cite{WOP11} and so-called supercollision cooling \cite{SRL12}. The latter is consistent with experimental observations \cite{GSR13,GSW13}. Recent calculations have also identified a remarkably fast rate of radiative transfer between graphene films \cite{IJJ12,RTD15}, graphene nanoribbons \cite{LZ15_2}, and extended heterostructures of graphene and hexagonal BN \cite{PLH16}, although all of them involve picosecond or even longer timescales. However, we need much faster transfer rates in order to prevent most of the electronic heat from being absorbed into phonons. We accomplish such a goal in this paper by resorting to graphene nanostructures capable of sustaining plasmons within an energy range that is commensurate with $\kB T$. Incidentally, radiative energy transfer from graphene electrons to optical phonons in a silica substrate has been argued to explain the measured saturation of conductivity in the carbon layer and provide a viable way of observing quantum friction \cite{VP11}.

Here we exploit the extraordinary optical and thermal properties of graphene to show that ultrafast radiative heat transfer can take place between neighboring nanoislands. The commonly accepted scheme for dissipation of the thermal energy produced by electronic and optical inelastic losses (i.e., energy transfer to valence and conduction electrons of the system, followed by relaxation into phonons and subsequent heat flow into the surrounding media) is here challenged by the radiative transfer mechanism taking place between neighboring structures within femtosecond timescales, thus overcoming electron relaxation into the atomic lattice. Using attainable graphene nanostructure designs, we find that ultrafast radiative heat transfer produces thermalization of two neighboring islands that results in $>$50\% of the electronic heat of the hot one being radiatively transferred to its colder neighbor. This extraordinary phenomenon is made possible by the large plasmonic field concentration that mediates the coupling between the neighboring graphene structures, as well as by the low specific electronic heat of this material \cite{paper235}. In particular, plasmons in this material exhibit unprecedentedly large electrical tunability accompanied by strong confinement of the measured fields \cite{paper196,FRA12}, which have recently enabled high mid-infrared sensitivity in the detection of proteins \cite{paper256} and other organic molecules \cite{HYZ16}. In a similar fashion, the ultrafast radiative heat transfer phenomenon here investigated can be actively switched on and off by gating the graphene structures.

\begin{figure*}
\begin{centering}
\includegraphics[width=0.55\textwidth]{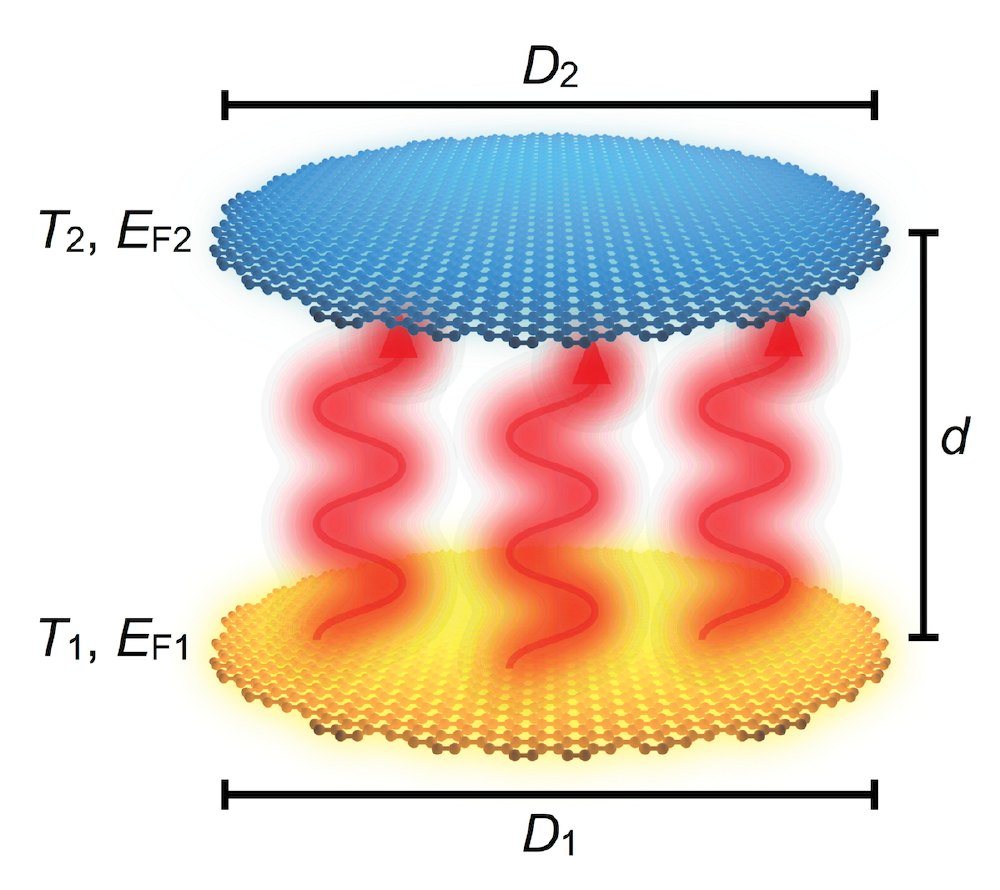}
\par\end{centering}
\caption{{\bf Sketch of the structure considered for ultrafast radiative heat transfer.} We study heat transfer between two parallel coaxial graphene disks placed in vacuum and separated by a small distance $d$. Each disk $\ell=1,2$ is characterized by its diameter $D_\ell$, Fermi energy $E_{{\rm F}\ell}$, and electron temperature $T_\ell$, with $T_1>T_2$.}
\label{Fig1}
\end{figure*}

\section{Results and Discussion}

\subsection{Radiative heat transfer between graphene nanodisks}

We focus on the system depicted in Fig.\ \ref{Fig1}, consisting of two parallel coaxial graphene nanodisks of diameters $D_{1}$ and $D_{2}$, separated by a distance $d$ between carbon planes, doped to Fermi levels $E_{\mathrm{F}1}$ and $E_{\mathrm{F}2}$, and having electronic temperatures $T_1>T_2$. For simplicity, we consider the disks to be placed in vacuum, as the conclusions of this work remain the same when the disks are surrounded by a dielectric material such as BN (e.g., $\epsilon\sim3.2$, see Fig.\ \ref{FigS1}). Heat is radiatively transferred from the hotter disk to the colder one as a result of thermal fluctuations in both disks, whose interaction is mediated by their self-consistent electromagnetic response. In fact, for the small size of the structures under consideration compared with the thermal wavelengths $\lambda_{T_\ell}$ (with $\ell=1,2$), retardation and magnetic response effects can be dismissed, so we only need to deal with charge fluctuations and their Coulomb interaction.

We calculate the heat transfer power (HTP) as the net balance of the work done by the thermally fluctuating charges of the hotter disk on the colder one minus the work done on the former by the fluctuating charges of the latter. This leads to a classical electromagnetic expression involving thermal fluctuations, which are evaluated by means of the fluctuation-dissipation theorem \cite{R1959_2,paper157}. A detailed self-contained derivation is offered in the Appendix, leading to a compact expression [Eq.\ (\ref{P1})] that is proportional to the integral over the exchanged frequency $\omega$. The integrand consists of the difference between the Bose-Einstein occupation numbers $n_\ell=\left[\exp(\hbar\omega/\kB T_\ell)-1\right]^{-1}$ of the two disks at their respective temperatures $T_\ell$, multiplied by a loss function that is determined by the disk susceptibilities $\chi_\ell$. The latter are dominated by plasmonic modes, which allow us to formulate a description in terms of plasmon wave functions (PWFs) \cite{paper228,paper257}. Only the lowest-order PWFs contribute significantly to the HTP for the range of geometrical parameters under consideration. Their explicit form (see Appendix), as well as full details on the PWF-based susceptibilities, are given in the Appendix. For coaxial disks (Fig.\ \ref{Fig1}), we find that modes of different azimuthal number $m$ do not mix, so we can separate their contributions to the HTP received by disk 2 as
\begin{widetext}
\begin{align}
P_2=\frac{2\hbar}{\pi}\sum_{m=0}^\infty(2-\delta_{m0})\int_0^\infty\omega\,d\omega
\,(n_1-n_2) \;{\rm Tr}\left[{\Delta^m}^\dagger\cdot v^m\cdot{\rm Im}\{\chi_1^m\}
\cdot v^m\cdot\Delta^m\cdot{\rm Im}\{\chi_2^m\}\right]
\label{P1disks}
\end{align}
\end{widetext}
(and also $P_1=-P_2$), where ${\rm Tr}[\dots]$ stands for the trace, the matrix $\Delta^m=(\mathbbm{I}-\chi_2^m\cdot v^m\cdot\chi_1^m\cdot v^m)^{-1}$ accounts for multiple scattering between the disks, $v^m$ describes their mutual Coulomb interaction, and $\mathbbm{I}$ is a unit matrix. The matrices $v^m$ and $\chi_\ell^m$ contain elements projected on the PWFs with $m$ azimuthal symmetry (see Appendix for detailed expressions). Incidentally, the leading $(2-\delta_{m0})$ factor reflects the fact that $m$ and $-m$ modes yield the same contribution.

In this formalism, the optical response of graphene is described through its surface conductivity $\sigma$, for which we adopt the local-RPA model \cite{paper235,GSC06,GSC09} [see Eq.\ (\ref{localRPA}) in the Appendix]. We remark that, besides the explicit dependence of $n_\ell$ on $T_\ell$, the temperature enters $\sigma$ through the chemical potential as well (see Appendix). It should be noted that, in contrast to extended graphene, the lack of translational invariance in nanostructures prevents us from using the full nonlocal RPA conductivity\cite{WSS06,HD07}. However, a full RPA description of the optical absorption of the system under consideration based on a previous implementation for finite structures \cite{paper183} reveals that nonlocal effects only play a small role (see Fig.\ \ref{FigS2}). We further analyze heat transfer between closely spaced extended graphene films, and more specifically, the contribution coming from parallel wave-vector components $\sim2\pi/D_\ell$, for which we find that nonlocal effects are also small for the graphene parameters under consideration (see Sec.\ \ref{SN4} and Fig.\ \ref{FigS4}), and therefore, we also expect them to be small for disks of diameter $D_\ell$.

Incidentally, as the HTP of Eq.\ (\ref{P1disks}) is an integrated quantity, it is not too sensitive to the model used for the graphene conductivity $\sigma$. This is corroborated in Fig.\ \ref{FigS3}(a,b), where we compare results obtained using either the local-RPA or the Drude model [Eq.\ (\ref{localRPA}) with the $E$ integral set to zero]. Only small discrepancies between the two models are observed at small separations $d$ in the resulting HTP. Actually, the small $d$ region is most sensitive to elements of the formalism such as the inclusion of multiple scattering in the optical response of the disks [$\Delta^m$ matrices in Eq.\ (\ref{P1disks}), see Fig.\ \ref{FigS5} for a comparison with results obtained by setting $\Delta^m=\mathbbm{I}$]. We also observe a mild dependence of the HTP on the value of the intrinsic electronic decay time (Fig.\ \ref{FigS6}), which we set to $\hbar\tau^{-1}=10\,$meV throughout this work. Additionally, we find good convergence of the HTP with the number of $m$'s and PWFs used in the calculations (Fig.\ \ref{FigS7}).

We stress that the relatively high temperatures under consideration (thousands of degrees) refer to the electronic gas of the material, which can be reached by optical pumping in the ultrafast regime \cite{LMS10,GPM13,NWG16}.

\begin{figure*}
\begin{centering}
\includegraphics[width=0.9\textwidth]{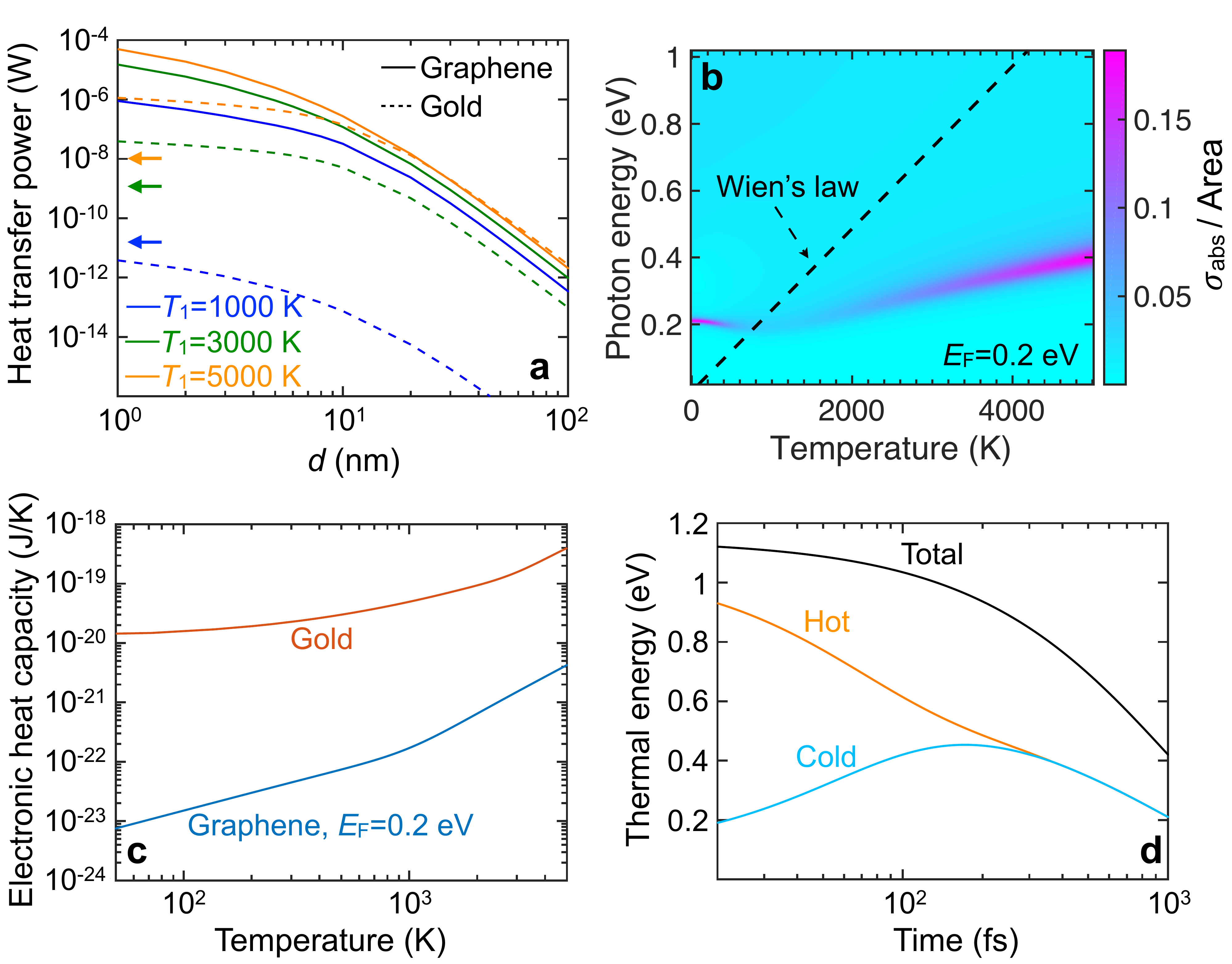}
\par\end{centering}
\caption{{\bf Thermal and optical properties associated with radiative heat transfer.} {\bf (a)} Dependence of the radiative heat transfer power (HTP) on the separation distance $d$ between two graphene nanodisks (solid curves) compared with two gold nanodisks (dashed curves, disk thickness $t=2\,$nm). All disks are 20\,nm in diameter. The HTP is plotted for different values of $T_1$ (see legend), while the cold disk is at ambient temperature $T_2=300\,$K. The arrows indicate the HTP between two blackbodies of an area equal to that of the present disks and placed at temperatures $T_1$ and $T_2$. Both graphene nanodisks are assumed to be doped with the same Fermi energy $E_{{\rm F}1}=E_{{\rm F}2}=0.2\,$eV and described by the local-RPA conductivity (see Appendix). {\bf (b)} Optical absorption cross-section $\sigma_{\rm abs}$ normalized to the graphene area for one of the graphene disks considered in (a) as a function of photon energy $\hbar\omega$ and temperature $T$. The dashed line corresponds to Wien's law, $\hbar\omega\approx2.82\,\kB T$. {\bf (c)} Temperature dependence of the electronic heat capacity for one of the graphene (blue curve, see Appendix) and gold (red curve, taken from Ref. \cite{LZC08}) nanodisks considered in (a). {\bf (d)} Illustrative example of the femtosecond dynamics of the electronic thermal energy in two graphene nanodisks under the conditions of (a) for a separation $d=1\,$nm, with initial temperatures $T_1=1000\,$K and $T_2=300\,$K. The electronic thermal energy is shown for both the initially hot (orange curve) and cold (cyan curve) nanodisks, as well as their sum (black curve).}
\label{Fig2}
\end{figure*}

The disk separation dependence of the HTP is studied in Fig.\ \ref{Fig2}(a) (solid curves) for 20\,nm graphene disks doped to a Fermi level $\EF=0.2\,$eV, with the hotter disk at different temperatures $T_1$ (see labels) and the colder one at room temperature $T_2=300\,$K. In general, higher temperatures $T_1$ lead to larger HTP, due in part to the $(n_1-n_2)$ factor in Eq.\ (\ref{P1disks}). At large separations $d\gg D_\ell$, only dipole-dipole interactions between the disks contribute efficiently to the transfer, leading to a $1/d^6$ dependence, in agreement with the asymptotic expression of Eq.\ (\ref{P1ana}) (see Appendix). A smooth convergence of the full calculation [Eq.\ (\ref{P1disks})] to this limit [Eq.\ (\ref{P1ana})] is observed in the additional calculations presented in Fig.\ \ref{FigS8}. The near-field character of heat transfer is further emphasized by considering the extension of the dominant dipole plasmon away from the disks (i.e., the electric-field amplitude decays by $1/e$ over a distance $\sim D/2\pi$, as estimated from the out-of-plane decay of plasmons in extended graphene for an equivalent wavelength $\sim D$), which explains the low slope in the curves of Fig.\ \ref{Fig2}(a) at small $d$'s.

As a reference, we compare these results with the HTP for gold disks of the same diameter [Fig.\ \ref{Fig2}(a), broken curves], which we describe through an effective surface conductivity obtained from the measured dielectric function \cite{JC1972} $\epsilon_{\rm Au}$ as $\sigma_{\rm Au}=\ii\omega t(1-\epsilon_{\rm Au})/4\pi$, where we take a thickness $t=2\,$nm. This approximation, which is reasonable because we are considering a small value of $t$ compared with the diameter (20\,nm), allows us to apply the same formalism as for graphene [Eq.\ (\ref{P1disks})]. Despite the larger thickness of the gold disks, their HTP is much smaller than for graphene. In fact, plasmons in the graphene disks lie in the mid-infrared region for the parameters under consideration (i.e., they energies are commensurate with $\kB T_1$), while those of the gold disks appear at much higher energies, and thus do not contribute efficiently to the heat transfer. This mismatch is partly alleviated at the highest temperature under consideration ($T_1=5000\,$K), for which gold and graphene disks exhibit similar HTPs in the large $d$ limit.

As an additional comparison, the left arrows in Fig.\ \ref{Fig2}(a) show an estimate obtained from the Stefan-Boltzmann law \cite{VP07} for radiative heat transfer between two blackbodies of an area equal to that of the present disks. As anticipated above, graphene outperforms blackbodies by several orders of magnitude.

The strength of their optical response influences the ability of the disks to transfer energy radiatively. This is examined in Fig.\ \ref{Fig2}(b), where we plot the absorption cross-section of one of the graphene disks considered in Fig.\ \ref{Fig2}(a). An intense plasmon feature is observed in the 0.2-0.4\,eV region, whose temperature dependence is inherited from the conductivity [Eq.\ (\ref{localRPA})]. The dashed line in Fig.\ \ref{Fig2}(b) shows the relation between the temperature and the photon energy according to Wien's law (i.e., the value of $\hbar\omega$ at the maximum of $\omega^3 n_\ell(\omega)$ as a function of $T_\ell$). This is relevant for the analysis of Eq.\ (\ref{P1disks}), in which a factor $\omega\,n_\ell(\omega)$ appears explicitly, whereas the remaining $\omega^2$ factor comes from the low $\omega$ limit of the ${\rm Im}\{\chi_\ell^m\}$ matrices [obviously, the full $\omega$ dependence of the integrand of Eq.\ (\ref{P1disks}) is more complex, as shown in Fig.\ \ref{FigS3}(g,h), but an analysis based on Wien's law is still informative]. Additionally, the response functions entering the trace in Eq.\ (\ref{P1disks}) display maxima near the plasmons, and therefore, the overlap between the dashed line and the plasmon in Fig.\ \ref{Fig2}(b) indicates that this excitation contributes efficiently to the HTP, thus providing a criterium for optimization. Incidentally, the plasmon dispersion and strength follow nonmonotonic behaviors resulting from the complex interplay between the increase in both the density of free charge carriers and the number of decay channels associated with single-electron transitions.

The electronic heat capacity provides a relation between the temperature and the amount of energy strored in the electron gas. In this respect, graphene is also advantageous relative to traditional plasmonic materials such as gold because its heat capacity is orders of magnitude smaller [Fig.\ \ref{Fig2}(c)] as a result of its conical band structure, in contrast to the parabolic dispersion of gold conduction electrons. In consequence, cooling the graphene electrons requires transferring a smaller amount of heat, thus making the process potentially faster. 

\subsection{Ultrafast radiative heat transfer regime}

We study the heat transfer dynamics by considering the electronic heat $Q_\ell$ deposited on each graphene disk $\ell$ and the evolution of these quantities according to the equations
\begin{align}
\dot{Q}_\ell=-\tau_{\rm ph}^{-1}Q_\ell+P_\ell,\;\;\;\;\;\;\;(\ell=1,2)
\label{eq:transfer}
\end{align}
where $P_\ell$ are the transfer powers given by Eq.\ (\ref{P1disks}), while $\tau_{\rm ph}$ is a phenomenological electron relaxation time (to phonons) that we approximate as 1\,ps, a value of the order of what is observed in pump-probe experiments \cite{DSC08,WOP11}. We note that the electronic heat of each disk $\ell$ depends on the electronic temperature $T_\ell$ as $Q_\ell=\beta\,\pi D_\ell^2(\kB T_\ell)^3/(2\hbar\vF)^2$ [see Eq.\ (\ref{Q1}) in the Appendix]. Also, the transfer powers $P_1$ and $P_2=-P_1$ [Eq.\ (\ref{P1disks})] implicitly depend on both temperatures $T_1$ and $T_2$. In order to make this clearer, we provide equations equivalent to Eqs.\ (\ref{eq:transfer}) at the end of the Appendix with a more explicit dependence on the temperatures, along with details of the numerical solution method. It should be pointed out that, because the electronic heat capacity in graphene is much smaller than that associated with the lattice, the temperature reached by the system when electrons and phonons are in thermal equilibrium is much smaller than the electron temperatures here considered after optical pumping. For this reason, we neglect the lattice in our analysis.

\begin{figure*}
\begin{centering}
\includegraphics[width=0.9\textwidth]{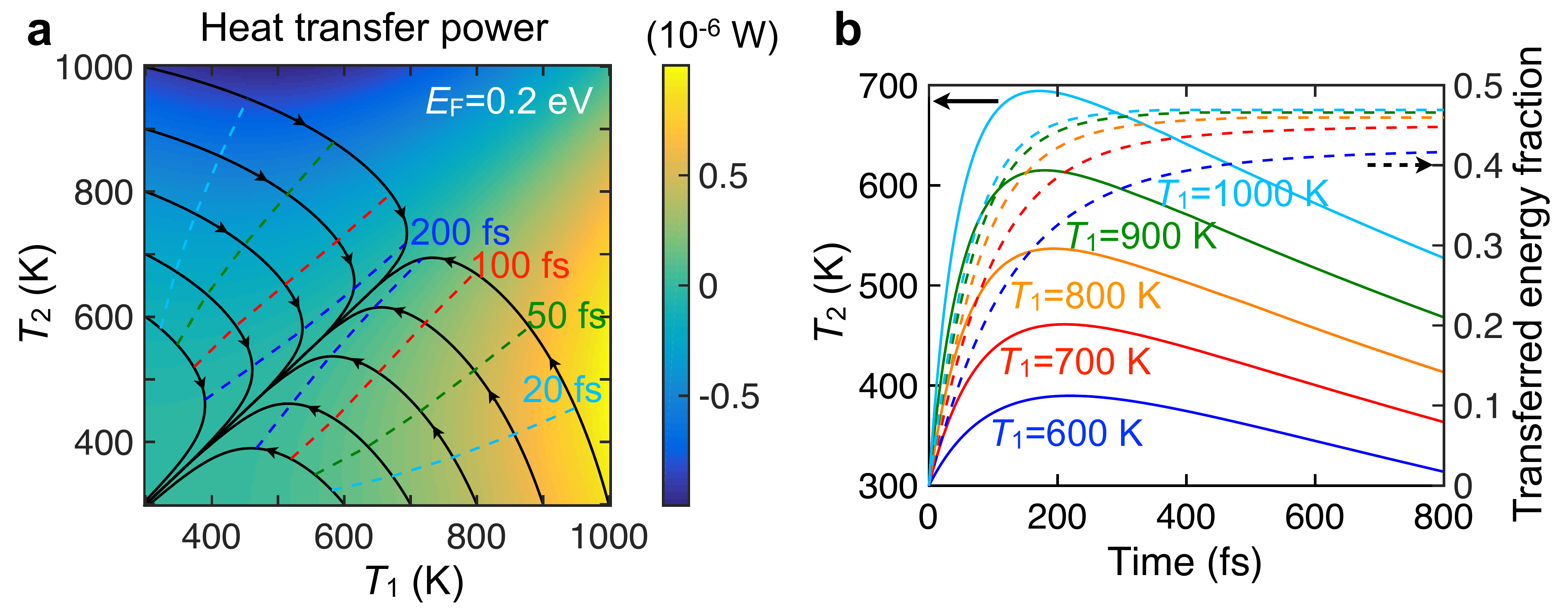}
\par\end{centering}
\caption{{\bf Temperature and temporal dependences of radiative heat transfer.} {\bf (a)} HTP between two graphene nanodisks (20\,nm diameter, $0.2\,$eV Fermi energy, $d=1\,$nm separation) as a function of $T_{1}$ and $T_{2}$ for the geometry of Fig.\ \ref{Fig1}. Black solid curves represent the evolution of the electron temperatures in the two nanodisks for different initial conditions. Dashed curves indicate the times (see labels) along the evolution of the solid curves from either the vertical or the horizontal axes of the plot. We assume an inelastic relaxation time (electron-lattice coupling) of 1\,ps . {\bf (b)} Temporal evolution of the electron temperature $T_{2}$ in the colder disk (initially at $T_{2}=300\,$K, left vertical scale, solid curves) and the transferred energy fraction from disk 1 to disk 2 (right vertical scale, dashed curves) for different initial electron temperatures of the hotter disk $T_{1}$ (see labels).}
\label{Fig3}
\end{figure*}

As an illustrative example, we show in Fig.\ \ref{Fig2}(d) the evolution of $Q_\ell$ according to Eqs.\ (\ref{eq:transfer}) for the two graphene disks considered in Fig.\ \ref{Fig2}(a) when they are prepared at initial temperatures $T_1=1000\,$K and $T_1=300\,$K: the cold disk more than doubles its electronic energy after $\sim200\,$fs of evolution (peak of cyan curve), when it has gained nearly the same amount of energy as the one dissipated to the atomic lattice (decay of black curve). Notably, the disks reach mutual thermal equilibrium after only $\sim250\,$fs, well before full relaxation takes place.

A more detailed study of the heat transfer dynamics is presented in Fig.\ \ref{Fig3} for 20\,nm graphene disks separated a distance of 1\,nm and doped to a Fermi energy of 0.2\,eV. The color plot of Fig.\ \ref{Fig3}(a) shows the HTP as a function of the temperatures in the two disks. Further calculations for a wider range of temperatures and more values of the disk diameters and the doping levels are presented in Figs.\ \ref{FigS9} and \ref{FigS10}. Obviously, the diagonal of this plot corresponds to zero transfer, when the two particles have the same temperature. The black solid curves represent the evolution of the disk temperatures starting from initial conditions at the plot axes (i.e., with one of the disks at 300\,K and the other one at higher temperature). The evolution is along the direction of the arrows, with positions at specific times indicated by the dashed curves. Interestingly, the evolution toward the diagonal (thermal equilibrium) is characterized by a significant increase in the temperature of the colder disk ($\Delta T\sim400\,$K) within the first 100-200\,fs, much faster than relaxation to the atomic lattice. This evolution involves the transfer of a large fraction of electronic heat to the colder disk, as shown in Fig.\ \ref{Fig3}(b): when the disks are prepared at 1000\,K and 300\,K initial temperatures, nearly 50\% of the electronic heat of the hot disk is transferred to the cold one within the first $\sim200\,$fs. We remark that fast transfers take place over a wide temperature range down to substantially smaller $T$'s [see Fig.\ \ref{Fig3}(a)]. These conclusions are maintained when considering larger disks (40\,nm diameter) or wider separations (3\,nm), as shown in Fig.\ \ref{FigS11}. They are also maintained when considering higher doping levels (Fig.\ \ref{FigS12}), well above the dipole plasmon energy, a condition for which nonlocal effects are particularly negligible. These supplementary figures also show that the results are robust with respect to variations in the disk diameters (e.g., similar conclusions are obtained for two dissimilar disks with diameters differing by a few nanometers).

\begin{figure*}
\begin{centering}
\includegraphics[width=0.9\textwidth]{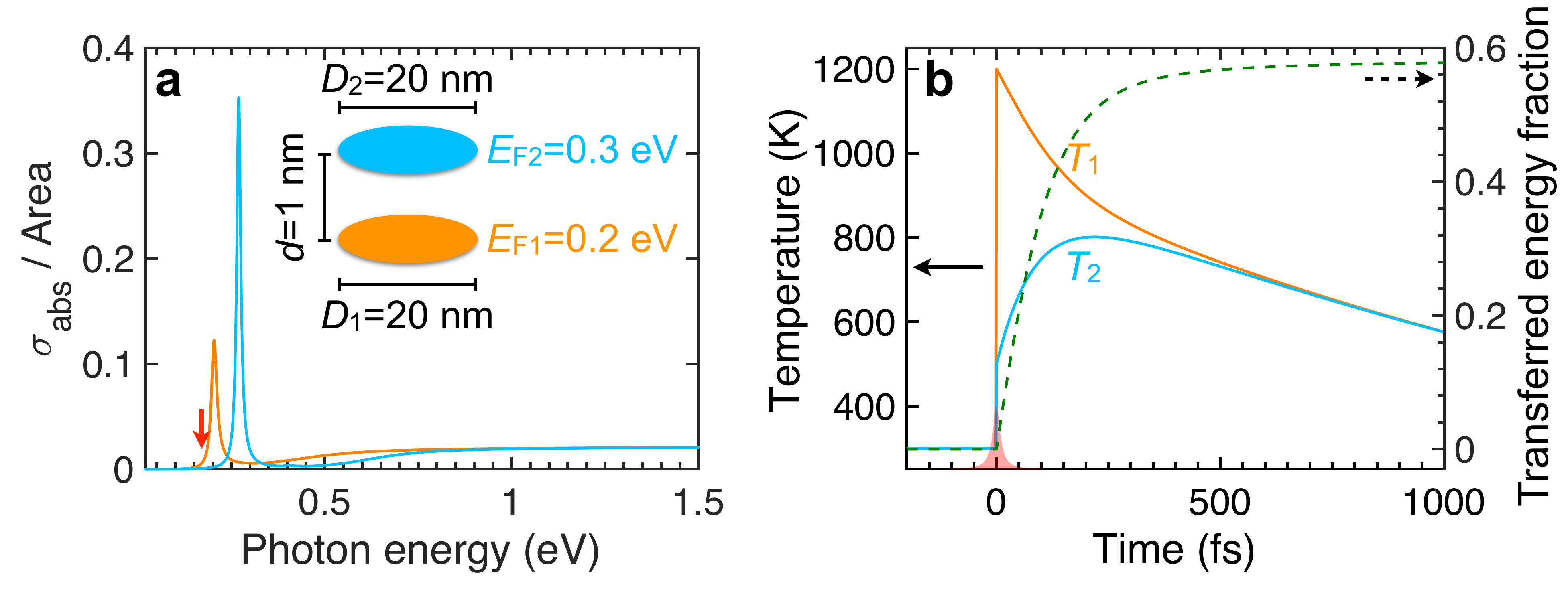}
\par\end{centering}
\caption{{\bf Ultrafast radiative heat transfer induced by optical pumping.} {\bf (a)} Normalized absorption cross-sections $\sigma_{\rm abs}$ of two graphene nanodisks (20\,nm diameter, $d=1\,$nm separation) at the same initial temperature of 300\,K but doped with different Fermi energies (see inset). {\bf (b)} Time evolution of the electron temperatures $T_{1}$ and $T_{2}$ (left vertical scale, solid curves) and transferred energy fraction (right vertical scale, dashed curve) after optical pumping ($150\,$mJ$\,$m$^{-2}$ light fluence, 0.17\,eV photon energy, as indicated by the red arrow of (a)).}
\label{Fig4}
\end{figure*}

In practical implementations, optical pumping with femtosecond laser pulses grants us access into the ultrafast regime, allowing us to reach high electron temperatures such as those considered in this work \cite{GSM11,TSJ13,MTP15}. Additionally, the amount of optically absorbed energy depends on the pump frequency relative to the plasmons of the system \cite{EBJ11}. This idea can be exploited to pump neighboring graphene disks in such a way that one of them absorbs much more energy than the other, just by tuning the pump laser near the plasmon of one of the disks and away from the plasmons of the other disk. We thus need disks of either different diameters or different Fermi levels. We consider the latter possibility, which can be realized in practice through the variation in intrinsic doping produced by an asymmetric dielectric environment, or also by creating different potential landscapes through an asymmetric doping geometry. The system under investigation is depicted in the inset of Fig.\ \ref{Fig4}(a): two 20\,nm graphene disks, separated by 1\,nm, initially placed at 300\,K, and doped to Fermi energies 0.2\,eV and 0.3\,eV, respectively. We consider optical pumping at a photon energy of 0.17\,eV with a fluence of $150\,$mJ$\,$m$^{-2}$. The pulse energy is closer to the lower doping disk [Fig.\ \ref{Fig4}(a)], and thus, this is the one that reaches a higher temperature. For simplicity, we assume instantaneous pumping (i.e., a $\delta$-function temporal profile of the pulse), which rapidly elevates the electron temperatures to $T_1\sim1200\,$K and $T_2\sim500\,$K [Fig.\ \ref{Fig4}(b), left end]. Interestingly, although the plasmons in the two disks are off-resonance before irradiation, optical pumping produces a larger blue shift in the hotter disk, bringing it on resonance with the initially bluer plasmon of the colder disk. Ultrafast radiative heat transfer is again observed, leading to mutual equilibrium between the disks ($T_1\approx T_2$) within $\sim500\,$fs, which is accompanied by nearly 60\% of the electronic heat of disk 1 being transferred to disk 2. We remark that higher that 50\% transferred energy fraction is made possible by the doping asymmetry, which directly affects the heat capacity (see Appendix). An interesting question for future studies relates to the maximum energy fraction that can be transferred in optimized structures.

\section{Concluding Remarks}

Our prediction of ultrafast radiative heat transfer in graphene provides a fundamentally unique scenario: radiative coupling is capable of evacuating electronic heat from a nanoisland to a surrounding structure fast enough to prevent substantial relaxation into the atomic lattice. This is accomplished with attainable geometrical and material parameters: tens of nanometers in lateral size $D$ in structures that can be patterned through state-of-the-art lithography \cite{HBE15,paper256} and bottom-up synthesis \cite{CRJ10,HTT13,YXL13}; vertical separations of a few nanometers, as provided by van der Waals atomic layer spacers \cite{WMH13,NMC16,MSV16}; tenths of electronvolts Fermi energy $\EF$, controllable through electrical gating \cite{LHJ08,CPB11}; and electron temperatures $T$ of thousands of degrees reached by ultrafast optical pumping \cite{LMS10,GPM13,BTM13,NWG16}.

Although we have focused on disks for computational convenience, we expect our conclusions to be maintained for other geometries of similar lateral size because the heat transfer power is a frequency-integrated quantity that should be qualitatively independent of the actual spectral position of the plasmon modes, as long as they overlap with Wien's law [see Fig.\ \ref{Fig2}(b)] and they are highly correlated with each other in the two islands. This correlation can be facilitated if the islands are nearly identical in shape and size. Actually, this is a condition that can be accomplished through lateral patterning of a stack formed by two graphene films and an atomically thin van der Waals layer spacer, using for example e-beam lithography.

In practice, the disks could have intrinsic doping due to interaction with a dielectric environment, which can change the Fermi energy by as much as $\sim0.3\,$eV. Obviously, because the disks do not have electrical connectivity, their control through electrostatic gating presents a challenge. However, gating should be possible in a configuration consisting of neighboring graphene ribbons, which can be biased and exposed to distant gates. The contacts can be placed far from the ribbon region in which heat transfer takes place, while the gates can also be 100s nm away and thus should not affect the heat transfer.

\begin{table*}
\begin{centering}
\begin{tabular}{|c|cccccc|}
\hline 
\backslashbox{$\nu$}{$m$} & 0 & 1 & 2 & 3 & 4 & 5 \tabularnewline
\hline 
1 & 0.0234 & 0.0720 & 0.0402 & 0.0283 & 0.0220 & 0.0181 \tabularnewline
2 & 0.0123 & 0.0165 & 0.0130 & 0.0109 & 0.0094 & 0.0083 \tabularnewline
3 & 0.0084 & 0.0101 & 0.0086 & 0.0076 & & \tabularnewline
4 & & 0.0073 & & & & \tabularnewline
\hline 
\end{tabular}
\par\end{centering}
\caption{{\bf Eigenvalues associated with the disk PWFs.} We list the values of $-\eta_{m\nu}$ corresponding to the disk PWFs $\rho_{m\nu}$ considered in Fig.\ \ref{Fig5} [see Eqs.\ (\ref{pwfdisk})].}
\label{Table1}
\end{table*}

Our choice of parameters leads to graphene plasmon energies \cite{paper235} $\hbar\omega_{m\nu}\sim e\sqrt{\EF/(-\pi\eta_{m\nu}D)}$ (as estimated from a Drude model description for the graphene conductivity, see Table\ \ref{Table1} for values of the eigenvalue $\eta_{m\nu}$ associated with disk plasmons) that are commensurate with $\kB T$ (i.e., they overlap the broad spectral peak of thermal emission, see Fig.\ \ref{FigS3}). As a consequence, the characteristic time interval $\tau_{\rm RHT}$ required to radiatively transfer a sizable fraction of the electronic heat energy is reduced to the femtosecond domain.

A simple dimensional analysis reveals that the HTP is proportional to $\EF/D$, provided the ratios of disk diameters and temperatures, as well as $d/D$ and the quantity $\EF/D T^2$, are kept constant (see also Figs.\ \ref{FigS9} and \ref{FigS10}). The optimum temperature at which maximum transfer takes place scales as $T\propto\sqrt{\EF/D}$. Additionally, we find the scaling $\tau_{\rm RHT}\propto\EF D^3$ with Fermi energy and lateral size, and therefore, low doping levels and small sizes enable faster cooling. These conclusions are consistent with the detailed numerical analysis of $\tau_{\rm RHT}$ presented in Fig.\ \ref{FigS13}.

We stress that the formalism developed in the Appendix can be readily applied to study radiative coupling assisted by fluctuations of other types of excitations besides plasmons, such as optical phonons in 2D polar materials, whose relative characteristic transfer time deserves further analysis.

Another interesting possibility consists in combining more than two structures. This could be used to accelerate the rate of heat evacuation and achieve greater control over the spatial flow of radiative heat transfer. Higher transfer rates could be also obtained through lateral shape optimization or by relying on other carbon allotropes such as carbon nanotubes. Additionally, similar fast transfers should be enabled by a wide range of existing atomic-scale materials capable of sustaining confined optical excitations \cite{BFG16} (e.g., exciton polaritons in dichalcogenides). Besides the fundamental interest of this line of research, electronic cooling via radiative heat transfer constitutes a promising avenue to effectively suppress relaxation to the atomic lattice, thus preventing thermal damage in nanoscale devices.

\appendix

\section{Theory of radiative heat transfer}

We consider two structures labeled by the index $\ell=1,\,2$, each of them assumed to be in internal thermal equilibrium at a temperature $T_\ell$. Radiative heat transfer can take place if $T_1\neq T_2$, mediated by electromagnetic interaction at characteristic frequencies $\sim\kB T_\ell/\hbar$ \cite{paper198}. We further assume the corresponding light wavelengths $\sim2\pi\hbar c/\kB T_\ell$ to be much smaller than the size of the structures. The response of the latter can be then described in the quasistatic limit through their susceptibilities $\chi_\ell(\rb,\rb',\omega)$, which are defined as the induced charge density distribution at $\rb$ produced by a unit potential point source oscillating with frequency $\omega$ at $\rb'$. The charge density induced in the $\ell$ structure by a monochromatic potential $\phi(\rb)\exp(-\ii\omega t)+{\rm c.c.}$ is then given by $\int d^3\rb'\chi_\ell(\rb,\rb',\omega)\phi(\rb')\exp(-\ii\omega t)+{\rm c.c.}$ Incidentally, although the emission of radiation away from the system is not accounted for within the quasistatic limit, this is a negligible contribution for the small structures under consideration, in which radiative heat transfer and relaxation to the atomic lattice occur at a much faster rate.

We express the net power received by structure 2 as the work $P_{2\leftarrow1}$ done on 2 by charges fluctuating in 1 minus the work $P_{1\leftarrow2}$ done on 1 by charges fluctuating in 2. It is enough to calculate the latter in detail, because the former is simply obtained by interchanging the subindices 1 and 2 in the resulting expression. We start from $P_{1\leftarrow2}=-\left\langle \int d^3\rb\,\jb_1(\rb,t)\cdot\nabla\phi_2(\rb,t)\right\rangle$, which is the work exerted by the electric field $-\nabla\phi_{2}(\rb,t)$ produced by fluctuations in 2, acting on the current $\jb_1(\rb,t)$ of 1. Here, $\langle\dots\rangle$ denotes the average over thermal fluctuations, the space integral extends over the entire 3D space, and the function $\jb_1$ is a distribution that vanishes outside the graphene and exhibits a singularity at the edge. Integrating the $\nabla$ operator by parts, writing the electric potential $\phi_2$ in terms of the charge $\rho_2$ {\it via} the Coulomb potential $v(\rb,\rb')$ (e.g., $v=1/\epsilon\,|\rb-\rb'|$ in a homogeneous medium of permittivity $\epsilon$), and using the continuity equation $\nabla\cdot\jb_1=-\partial_t\rho_1$, we find $P_{1\leftarrow2}=-\left\langle \int d^3\rb d^3\rb'\,\partial_t\left(\rho_1(\rb,t)\right)v(\rb,\rb')
\rho_{2}(\rb',t)\right\rangle$, or equivalently,
\begin{widetext}
\begin{align}
P_{1\leftarrow2}
&=\ii\iint\frac{d\omega d\omega'}{(2\pi)^2}
\,\omega\, \ee^{-\ii(\omega+\omega')t}
\left\langle \int d^3\rb d^3\rb'\,\rho_1(\rb,\omega)v(\rb,\rb')
\rho_2(\rb',\omega')\right\rangle
\nonumber \\
&=\ii\iint\frac{d\omega d\omega'}{(2\pi)^2}
\,\omega\,
\ee^{-\ii(\omega+\omega')t}
\left\langle\rho_1(\omega)^{\rm T}\cdot v\cdot\rho_2(\omega')\right\rangle,
\label{P12a}
\end{align}
\end{widetext}
where we have expressed the charges in frequency space $\omega$ and replaced $\partial_t$ by $-\ii\omega$. The last line of Eq.\ (\ref{P12a}) implicitly defines a matrix notation in which $\rb$ and $\rb'$ are used as matrix indices, while the dot indicates matrix multiplication. In this notation, $\rho_\ell$ are column vectors, $v$ and $\chi_\ell$ are matrices, and $\rho_\ell^{\rm T}$ is the transpose of $\rho_\ell$.

The self-consistent charges $\rho_\ell$ produced by the fluctuating charge $\rho_2^{\rm fl}$ are now obtained from the relations
\begin{align}
\rho_1&=\chi_1\cdot v\cdot\rho_2, \nonumber\\
\rho_2&=\chi_2\cdot v\cdot\rho_1+\rho_2^{\rm fl}, \nonumber
\end{align}
where we work in the frequency domain and use the matrix notation introduced above. We remark that $\rho_2^{\rm fl}(\rb,\omega)$ vanishes for $\rb$ outside structure 2, while $\chi_\ell(\rb,\rb',\omega)$ vanishes for $\rb$ or $\rb'$ outside $\ell$. By construction, $v(\rb,\rb')$ only needs to be evaluated for $\rb$ and $\rb'$ sitting at different structures. Inserting the solution of these equations into Eq.\ (\ref{P12a}), we find
\begin{widetext}
\begin{align}
P_{1\leftarrow2}=\ii&\iint\frac{d\omega d\omega'}{(2\pi)^2}
\,\omega\,
\ee^{-\ii(\omega+\omega')t}
\label{P12b}\\
&\times\;\int d^3\rb\int d^3\rb'
\left\langle
\left[\chi_1(\omega)\cdot v\cdot
\Delta(\omega)\cdot\rho_2^{\rm fl}(\omega)\right]\big|_\rb
\,v(\rb,\rb')\,
\left[\Delta(\omega')
\cdot\rho_2^{\rm fl}(\omega')\right]\big|_{\rb'}
\right\rangle,
\nonumber
\end{align}
where
\begin{align}
\Delta=(\mathbbm{I}-\chi_2\cdot v\cdot\chi_1\cdot v)^{-1},
\label{Delta}
\end{align}
whereas $\mathbbm{I}$ is the unit matrix (i.e., $\delta(\rb-\rb')$). Now, the average over thermal fluctuations can be carried out using the fluctuation-dissipation theorem \cite{N1928,CW1951,R1959_2,paper157}
\begin{align}
\left\langle \rho_\ell^{\rm fl}(\rb,\omega)\rho_{\ell'}^{\rm fl}(\rb',\omega')\right\rangle =-4\pi\hbar\delta_{\ell\ell'}\,\delta(\omega+\omega')
\;\left[n_\ell(\omega)+1/2\right]
\;{\rm Im}\left\{\chi_\ell(\rb,\rb',\omega)\right\},
\label{FDT}
\end{align}
where $n_\ell(\omega)=\left[\exp(\hbar\omega/\kB T_\ell)-1\right]^{-1}$
is the Bose-Einstein distribution at temperature $T_\ell$ (i.e., for structure $\ell$). A detailed self-contained derivation of Eq.\ (\ref{FDT}) is offered in Sec.\ \ref{SN1}. We find Eq.\ (\ref{P12b}) to reduce to
\begin{align}
P_{1\leftarrow2}
=\frac{2\hbar}{\pi}\int_0^\infty\omega\,d\omega
\,(n_2+1/2) \;{\rm Tr}\left[
\Delta^\dagger\cdot v\cdot{\rm Im}\{\chi_1\}\cdot v\cdot \Delta\cdot{\rm Im}\{\chi_2\}
\right],
\label{P12c}
\end{align}
where ${\rm Tr}[\dots]$ stands for the trace, $\dagger$ refers to the conjugate transpose, and a dependence on $\omega$ is understood in all quatities. In the derivation of Eq.\ (\ref{P12c}), we have used the properties $v=v^{\rm T}$ and $\chi_\ell=\chi_\ell^{\rm T}$ (reciprocity), $\chi_\ell(\omega)=\chi_\ell^*(-\omega)$ (causality), $[n_\ell(\omega)+1/2]=-[n_\ell(-\omega)+1/2]$, ${\rm Tr}[A]={\rm Tr}[A^{\rm T}]$, and ${\rm Tr}[A\cdot B]={\rm Tr}[B\cdot A]$ (see Sec.\ \ref{SN2} for further details).

Finally, the net power received by 2 is obtained from
\begin{align}
P_2&=P_{2\leftarrow1}-P_{1\leftarrow2} \nonumber\\
&=\frac{2\hbar}{\pi}\int_0^\infty\omega\,d\omega
\,(n_1-n_2) \;{\rm Tr}\left[\Delta^\dagger\cdot v\cdot{\rm Im}\{\chi_1\}
\cdot v\cdot\Delta\cdot{\rm Im}\{\chi_2\}\right],
\label{P1}
\end{align}
\end{widetext}
where the matrix $\Delta$ [see Eq.\ (\ref{Delta})] accounts for multiple scattering between the two structures. Incidentally, the latter cannot be ignored at short separations, as shown in Fig.\ \ref{FigS5}. From the invariance of the expression in the square brackets of Eq.\ (\ref{P1}) under exchange of the subindices 1 and 2 (see Sec.\ \ref{SN2}), we confirm the expected result $P_1=-P_2$.

Finally, for structures separated by a large distance $d$ compared to their sizes, in virtue of induced-charge neutrality (i.e., $\int d^3\rb\,\chi_\ell(\rb,\rb',\omega)=0$ for each $\ell$), the leading contribution to $v$ is the dipole-dipole interaction. For parallel disks placed in vacuum, like the ones considered throughout this work, neglecting multiple scattering (i.e., taking $\Delta=\mathbbm{I}$), we find from Eq.\ (\ref{P1})
\begin{align}
P_2\approx\frac{4\hbar}{\pi d^{6}}\int_0^\infty\omega\, d\omega\;(n_1-n_2)\;{\rm Im}\{\alpha_1\}{\rm Im}\;\{\alpha_2\},
\label{P1ana}
\end{align}
where
\begin{align}
\alpha_\ell(\omega)=-\int x\,d^3\rb \int x'\,d^3\rb' \chi_\ell(\rb,\rb',\omega)
\label{alpha0}
\end{align}
is the polarizability of disk $\ell$ along a direction $x$ parallel to it. An extra factor of 2 has been introduced in Eq.\ (\ref{P1ana}) to account for the two equivalent orthogonal directions in the planes of the disks. The convergence of Eq.\ (\ref{P1}) toward Eq.\ (\ref{P1ana}) is illustrated by calculations presented in Fig.\ \ref{FigS8}.

\section{Description of graphene islands through plasmon wave functions (PWFs)}

We now apply the above formalism to two parallel graphene islands placed in a homogeneous medium of permittivity $\epsilon$ and separated by a vertical distance $d=|z_\ell-z_{\ell'}|$ along their normal direction $z$. It is then convenient to use an eigenmode expansion for the response of each island $\ell$ \cite{paper228,paper257}. This allows us to define a complete set of PWFs $\rho_{\ell j}$ and real eigenvalues $\eta_{\ell j}$, where $j$ is a mode index. More precisely, the susceptibility of the $\ell$ island, taking to be in the $z=z_\ell$ plane, admits the rigorous exact expansion \cite{paper228}
\begin{align}
\chi_\ell(\rb,\rb',\omega)=\frac{\epsilon}{D_\ell^3}\sum_{j}\frac{\rho_{\ell j}(\th)\rho_{\ell j}(\th')}{1/\eta_{\ell j}-1/\eta^{(\ell)}(\omega)}\;\delta(z-z_\ell)\delta(z'-z_\ell),
\label{chil}
\end{align}
where $j$ runs over eigenmodes, we use the notation $\rb=(D_\ell\th,z)$, $\th$ is an in-plane coordinate vector normalized to a characteristic length of the structure $D_\ell$ (we use the diameter for disks), and
\begin{align}
\eta^{(\ell)}(\omega)=\frac{\ii\sigma_\ell(\omega)}{\epsilon\,\omega D_\ell}
\label{eta0}
\end{align}
incorporates the response of the graphene through its local conductivity $\sigma_\ell(\omega)$. It should be noted that the latter depends on $\ell$ {\it via} the level of doping and the temperature (see below). The PWFs and their eigenvalues satisfy the orthogonality relation \cite{paper228}
\begin{align}
\int d^2\th \int d^2\th'\;
\frac{\rho_{\ell j}(\th) \rho_{\ell j'}(\th')}{|\th-\th'|}=-\frac{\delta_{jj'}}{\eta_{\ell j}}.
\label{ortho}
\end{align}
For islands with the same geometrical shape (e.g., disks), the PWFs and eigenvalues are independent of size $D_\ell$, even if $D_1\neq D_2$.

We can readily use Eq.\ (\ref{chil}) to evaluate the heat transfer rate according to Eq.\ (\ref{P1}). With some straightforward redefinitions, these equations remain the same, but now the coefficients of the matrices that they contain are labeled by eigenmode indices $j$ instead of spatial coordinates $\rb$. More precisely, $\chi_\ell$ becomes a diagonal matrix of coefficients
\begin{align}
\chi_{\ell,jj'}=\delta_{jj'}\;\frac{\epsilon}{D_\ell^3}\frac{1}{1/\eta_{\ell j}-1/\eta^{(\ell)}},
\nonumber
\end{align}
while the matrix elements of the Coulomb interaction reduce to
\begin{align}
v_{jj'}=\frac{D_\ell^2D_{\ell'}^2}{\epsilon}\int d^2\th\int d^2\th'\;
\frac{\rho_{\ell j}(\th)\rho_{\ell' j'}(\th')}{\sqrt{|D_\ell\th-D_{\ell'}\th'|^2+d^2}}
\label{vjjp}
\end{align}
when the operators to the left and right of $v$ are referred to islands $\ell$ and $\ell'$, respectively. Incidentally, in this work we focus on disk dimers that share the same axis of symmetry; an eventual lateral displacement ${\bf b}$ between the islands is however easy to implement by adding it to $D_\ell\th-D_{\ell'}\th'$ in the above expression.

In this PWF formalism, inserting Eq.\ (\ref{chil}) into Eq.\ (\ref{alpha0}), we find that the polarizability of a graphene island along a given in-plane symmetry direction $x$ is given by
\begin{equation}
\alpha_\ell(\omega)=\epsilon D_\ell^3\sum_j\frac{\zeta_j^2}{1/\eta^{(\ell)}-1/\eta_j},\label{alpha}
\end{equation}
where $\zeta_j=\int \theta_x\,d^2\th \rho_j(\th)$ is a normalized plasmon dipole moment.

\section{PWFs for disks}

In the disk geometry, the azimuthal number $m$ provides a natural way of classifying the PWFs. More precisely, we can label them using a double index $(m\nu)$ and separate the radial and azimuthal dependences as
\begin{subequations}
\begin{equation}
\rho_{m\nu}^{\rm c}(\th)=\rho_{m\nu}(\theta)\cos(m\varphi_{\th}),\;\;\;\;(m\ge0),
\end{equation}
\begin{equation}
\rho_{m\nu}^{\rm s}(\th)=\rho_{m\nu}(\theta)\sin(m\varphi_{\th}),\;\;\;\;(m\ge1).
\end{equation}
\label{pwfdisk}
\end{subequations}
\noindent We insist that these PWFs are the same for both disks in a dimer, as they are independent of disk size, and therefore, we drop the disk index $\ell$ for them. We also note that the PWFs are doubly degenerate for $m>0$ (i.e., they share the same eigenvalue $\eta_{m\nu}$ and radial component $\rho_{m\nu}(\theta)$ for both sine and cosine azimuthal dependences). We obtain the radial component $\rho_{m\nu}(\theta)$ by solving the Maxwell equations numerically using the boundary-element method \cite{paper040} (BEM) for a self-standing disk of small thickness $t\sim D/100$ compared with its diameter $D$. The disk is described by a dielectric function $\epsilon=1+4\pi\ii\sigma/\omega t$, where $\sigma$ is the Drude graphene conductivity (the actual model used for $\sigma$ is irrelevant, as the PWFs depend only on geometry and not on the specifics of the material). In the limit of small damping, the plasmons emerge as sharp, spectrally-isolated features in the local density of optical states (LDOS) \cite{paper176}. We average the LDOS over a set of off-center locations in order to access different $m$'s efficiently. The radial components of the PWFs are then retrieved from the induced charge density, while the eigenvalues are derived from the resonance condition $\eta_{m\nu}={\rm Re}\{\ii\sigma/\omega D\}$ at the corresponding LDOS peak maximum.

\begin{table*}
\begin{centering}
\begin{tabular}{|c|cc|ccc|cc|cc|c|c|}
\cline{2-12} 
\multicolumn{1}{c|}{} 
   &  \multicolumn{2}{c|}{$m=0$} 
   &  \multicolumn{3}{c|}{$m=1$}  
   &  \multicolumn{2}{c|}{$m=2$}   
   &  \multicolumn{2}{c|}{$m=3$} 
   &  $m=4$ & $m=5$ 
   \tabularnewline 
\hline 
\backslashbox{$\nu$}{$\nu'$} & 1               & 2         & 1          & 2          & 3         & 1        & 2          & 1        & 2          & 1        & 1        \tabularnewline \hline 
2 & 0.008        & 1         & 0.055   & 1         &             & 0.058 & 1        &   0.061 & 1         &  0.063 & 0.064 \tabularnewline
3 & 0.006        &  0.010 & 0.114   & -0.031 & 1          & 0.113 & -0.028 & 0.114 & -0.026 &           &            \tabularnewline
4 &                  &            & 0.078   & -0.019 & -0.026 &            &            &           &             &           &           \tabularnewline
\hline 
\end{tabular}
\par\end{centering}
\caption{{\bf Orthogonality of the disk PWFs.} Each entry in this table is obtained by numerically integrating the left-hand side of Eq.\ (\ref{ortho2}). The values of $m$, $\nu$, and $\nu'$ cover the ranges considered in Fig.\ \ref{Fig5} and Table\ \ref{Table1}. All diagonal entries ($\nu=\nu'$) are 1 by construction. We only show $\nu\ge\nu'$ values because the results are invariant under exchange of these two indices.}
\label{Table2}
\end{table*}

\begin{figure*}
\begin{centering}
\includegraphics[width=0.85\textwidth]{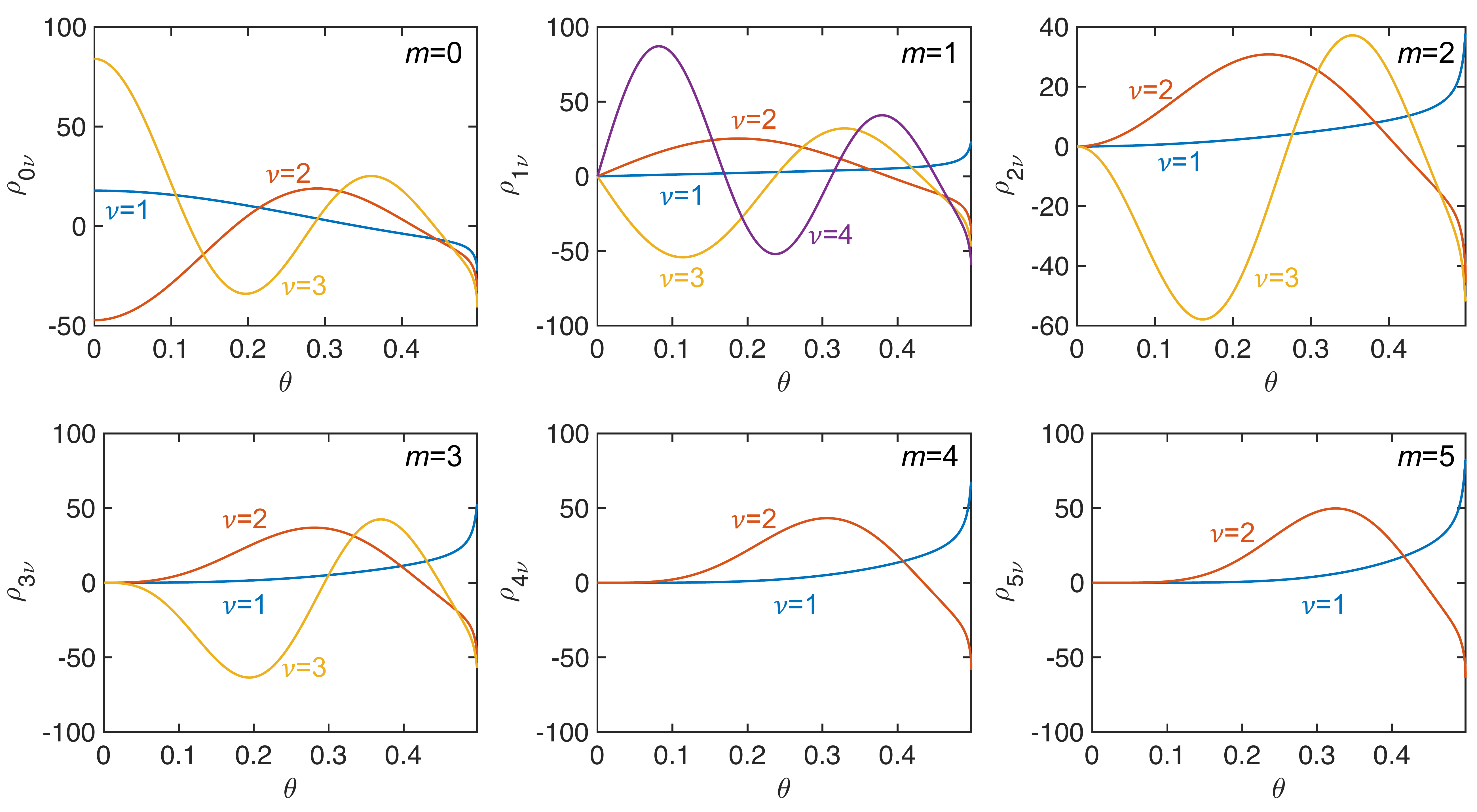}
\par\end{centering}
\caption{{\bf Radial components of the disk PWFs.} We show $\rho_{m\nu}(\theta)$ as defined in Eqs.\ (\ref{pwfdisk}) for several low values of $m$ and $\nu$ (see also Table\ \ref{Table1}).}
\label{Fig5}
\end{figure*}

By construction, $\rho_{m\nu}^{\rm c}$ and $\rho_{m\nu}^{\rm s}$ [see Eqs.\ (\ref{pwfdisk})] are mutually orthogonal according to Eq.\ (\ref{ortho}). Additionally, PWFs with different $m$'s are automatically orthogonal. For the remaining pairs of wave functions that share both the value of $m$ and the azimuthal dependence (either sine or cosine), Eq.\ (\ref{ortho}) reduces to
\begin{widetext}
\begin{align}
-4\pi\sqrt{\eta_{m\nu}\eta_{m\nu'}}\int_0^{1/2} \theta\,d\theta\,\rho_{m\nu}(\theta)\;
\int_0^{1/2} \theta'd\theta'\,\rho_{m\nu'}(\theta')\;
\int_0^\pi d\varphi\;
\frac{\cos(m\varphi)-(1/2)\delta_{m,0}}{\sqrt{\theta^2+{\theta'}^2-2\theta\theta'\cos\varphi}}=\delta_{\nu\nu'}.
\label{ortho2}
\end{align}
\end{widetext}
Our calculated radial PWFs, already normalized according to Eq.\ (\ref{ortho2}), are shown in Fig.\ \ref{Fig5} for the lowest values of $(m\nu)$, while their associated eigenvalues are given in Table\ \ref{Table1}. The orthogonality for $\nu\neq\nu'$ is rather satisfactory, as illustrated in Table\ \ref{Table2}, which shows the values obtained by numerically evaluating the left-hand side of Eq.\ (\ref{ortho2}).

Upon insertion of the disk PWFs in Eq.\ (\ref{vjjp}), we find that $v_{jj'}$ is diagonal by blocks (two blocks per $m$,  corresponding to the two different azimuthal symmetries of Eqs.\ (\ref{pwfdisk}) and each of them contributing the same to the HTP). As $\chi_{\ell,jj'}$ is diagonal, this allows us to write $P_2$ as a sum over $m$'s, essentially reflecting the fact that only modes of the same symmetry undergo mutual Coulomb interaction. The integrand of Eq.\ (\ref{P1}) then becomes an analytical function (see expressions for $n_\ell$, $\chi_\ell$, and $\Delta$ above), except for the integral over radial wave functions in $v_{jj'}$, for which we derive a computationally convenient expression in Sec.\ \ref{SN3}. We finally write Eq.\ (\ref{P1disks}) for the HTP, where the explicit dependence of the involved matrices on $m$ is indicated.

Only $m=1$ PWFs exhibit nonzero dipole moments $\zeta_\nu$ contributing to the polarizability $\alpha_\ell$ in Eq.\ (\ref{alpha}). More precisely, $\zeta_\nu$ is 0.84, 0.40, 0.11, and 0.08 for $\nu=1-4$, respectively. We use these coefficients and Eq.\ (\ref{alpha}) to obtain the absorption cross-section [Figs.\ 2(b), 4(a), and \ref{FigS3}(c-f)] as 
\begin{align}
\sigma_\ell^{\rm abs}(\omega)=(4\pi\omega/c){\rm Im}\{\alpha_\ell\}-(8\pi\omega^4/3c^4)|\alpha_\ell|^2,
\label{sigmaabs}
\end{align}
where the second term ($\propto|\alpha_\ell|^2$) is negligible for the small diameters of the disks under consideration ($\ll$\,light wavelength).

\begin{figure*}
\begin{centering}
\includegraphics[width=0.55\textwidth]{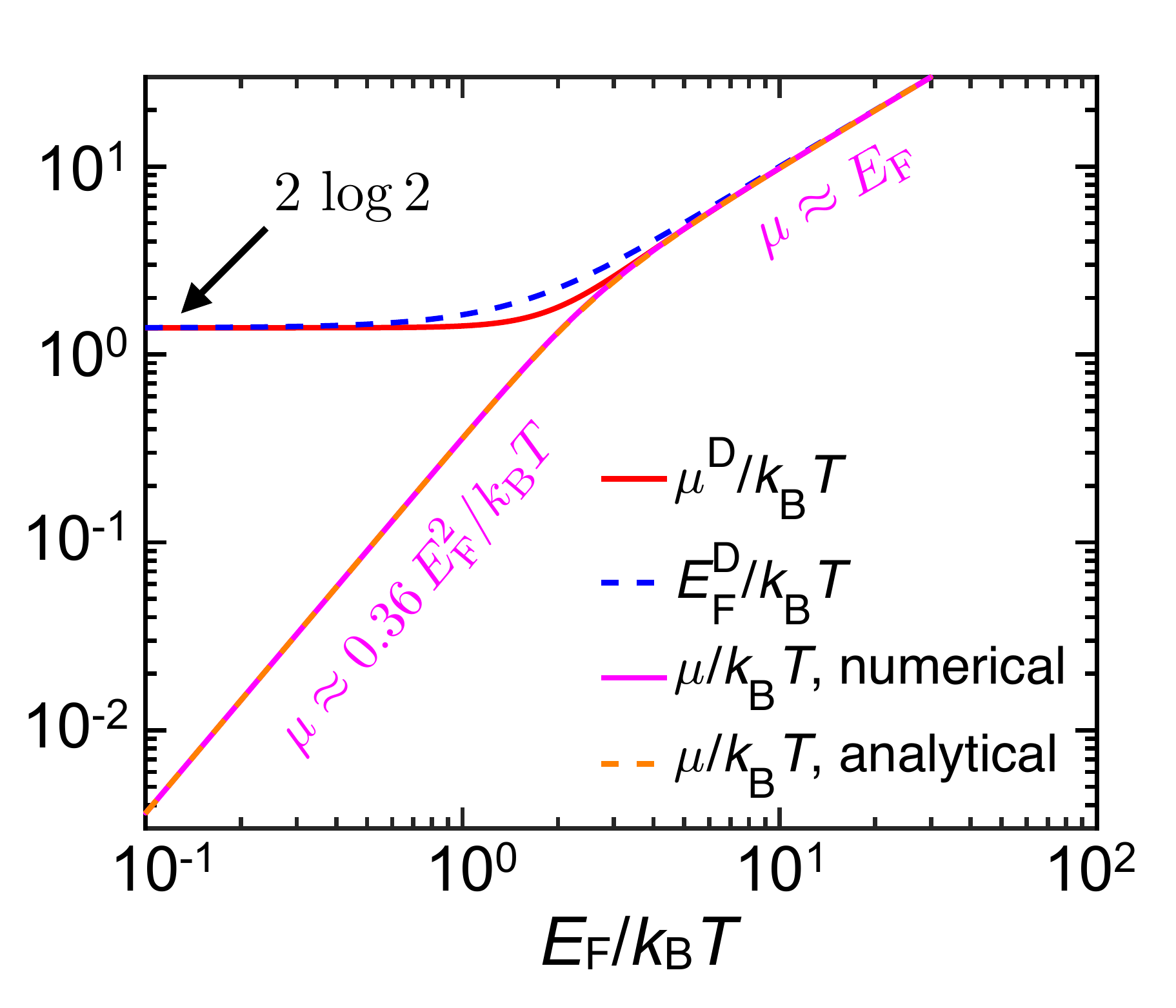}
\par\end{centering}
\caption{{\bf Graphene chemical potential and Drude weight.} We show the relation between the chemical potential $\mu$ and the Fermi energy $\EF$ in graphene, both of them normalized to $\kB T$. The direct numerical solution of Eq.\ (\ref{n1}) (pink solid curve) is nearly indistinguishable from the analytical expression of Eq.\ (\ref{n2}) (dashed orange curve). For completeness, we also plot the normalized Drude weight $\mu^{\rm D}/\kB T$ [red solid curve, see Eq.\ (\ref{muD})] and an approximate Drude weight $E_{\rm F}^{\rm D}/\kB T$ (dashed blue curve).}
\label{Fig6}
\end{figure*}

\section{Temperature-dependent graphene chemical potential}

At zero temperature, the Fermi energy $\EF$ describes a charge-carrier doping density $n$ subject to the relation \cite{CGP09} $\EF=\hbar\vF\sqrt{\pi n}$, where $\vF\approx10^6\,$m\,s$^{-1}$ is the Fermi velocity. This expression assumes a conical electronic band structure, which provides an accurate description for electron energies $E$ up to a couple of electronvolts away from the Dirac point \cite{BOS07}. For concreteness, we consider doping with electrons, as exactly the same results are obtained when doping with holes within the conical band approximation. At finite temperature $T$, the population of electronic states is given by the Fermi-Dirac distribution
\begin{align}
f_T(E)=\frac{1}{\ee^{(E-\mu)/\kB T}+1},
\nonumber
\end{align}
where $\mu$ is the chemical potential. The latter depends on temperature in such a way that the electron density
\begin{align}
n=\frac{4}{A}\sum_{\kb_\parallel}\left[f_T(E)+f_T(-E)-1\right]
\label{n0}
\end{align}
is maintained constant. Here, $A$ is the graphene area, the factor of 4 originates in valley and spin degeneracies, $\kb_\parallel$ is the parallel wave vector, $E=\hbar\vF k_\parallel>0$ is the electron energy in the upper Dirac cone, $f_T(E)$ is the electron population in that cone, and $1-f_T(-E)$ is the hole distribution in the lower cone. Recasting the sum over $\kb_\parallel$ into an integral (i.e., $\sum_{\kb_\parallel}\rightarrow(A/2\pi)\int_0^\infty k_\parallel dk_\parallel$), and defining $x=\hbar\vF k_\parallel/\kB T$, Eq.\ (\ref{n0}) becomes
\begin{align}
\left(\frac{\EF}{\kB T}\right)^2& \label{n1}\\
=& 2\int_0^\infty x dx\;\left[\frac{1}{\ee^{x-\mu/\kB T}+1}-\frac{1}{\ee^{x+\mu/\kB T}+1}\right].
\nonumber
\end{align}
Direct numerical integration of Eq.\ (\ref{n1}) allows us to obtain $\EF/\kB T$ as a function of $\mu/\kB T$. The result is plotted as a pink solid curve in Fig.\ \ref{Fig6}. Additionally, the large and small asymptotic $T$ limits of Eq.\ (\ref{n1}) (see pink labels in Fig.\ \ref{Fig6}) suggest the following approximate relation
\begin{equation}
\left(\frac{\EF}{\kB T}\right)^4=\left(\log^2 16\right)\left(\frac{\mu}{\kB T}\right)^2+\left(\frac{\mu}{\kB T}\right)^4,
\label{n2}
\end{equation}
which is in excellent agreement with the full solution of Eq.\ (\ref{n1}) (cf. pink-solid and dashed-orange curves in Fig.\ \ref{Fig6}). Also note that approximate\cite{A06,JKW16,STZ14} and asymptotic\cite{FLS14,MBT09} values for the Drude weight have been proposed to work well in different limits, although they lack the universal accuracy of Eq.\ (\ref{n2}).

\begin{figure*}
\begin{centering}
\includegraphics[width=0.55\textwidth]{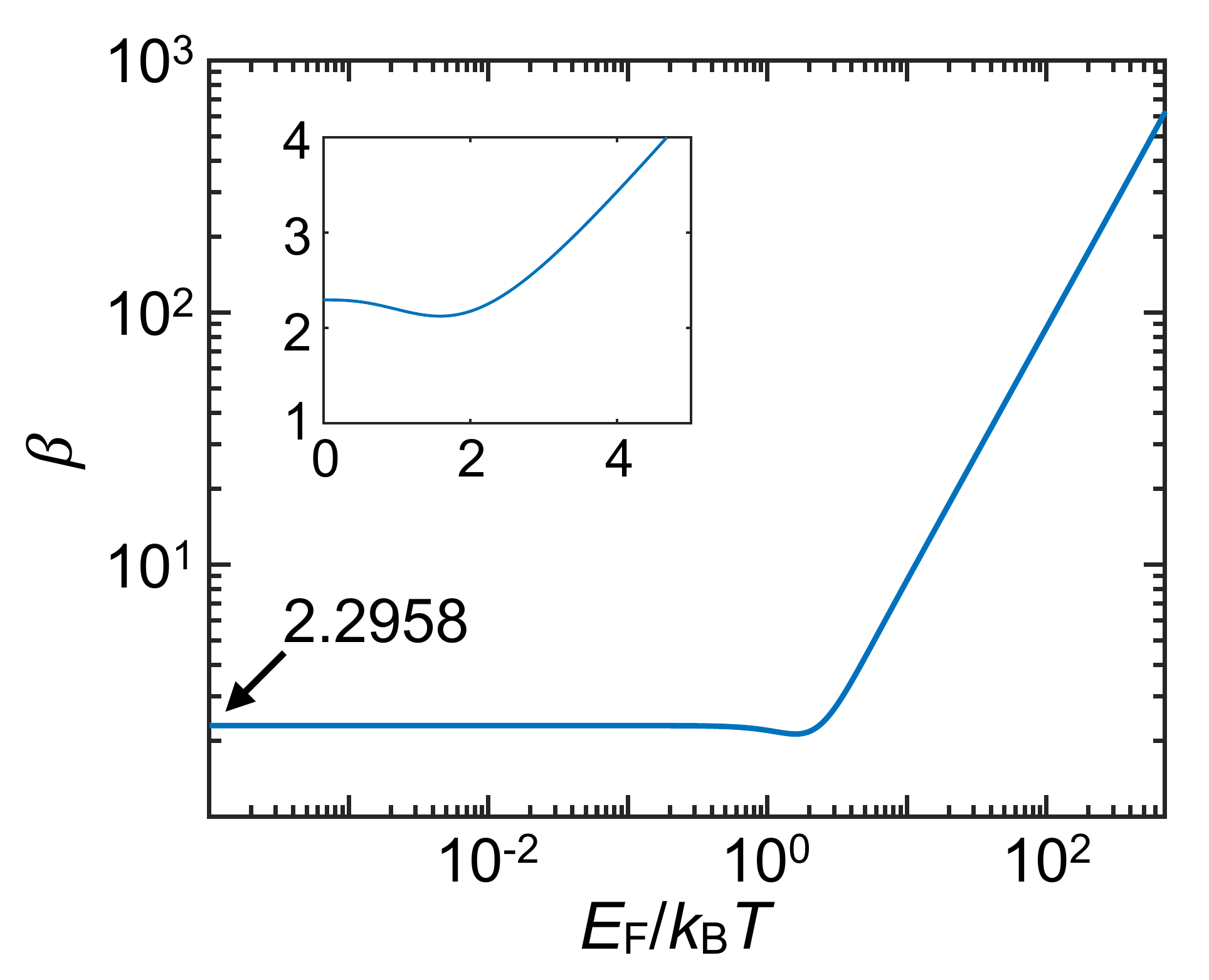}
\par\end{centering}
\caption{{\bf Graphene electronic heat.} We show the dependence of the thermal coefficient $\beta$ on $\EF/\kB T$, as calculated from Eq.\ (\ref{beta}). This parameter permits obtaining the electronic heat per unit of graphene area as $\beta\,(\kB T)^3/(\hbar\vF)^2$ [Eq.\ (\ref{Q1})]. The inset shows $\beta$ in linear scale.}
\label{Fig7}
\end{figure*}

\section{Electronic heat capacity of graphene}

The heat capacity is needed to relate the electronic thermal energy $Q$ to the electronic temperature $T$. By analogy to Eq.\ (\ref{n0}), the surface density of electronic thermal energy can be calculated as 
\begin{align}
\frac{Q}{A}=\frac{4}{A}\sum_{\kb_\parallel}E 
\big\{&[f_T(E)-\theta(\EF-E)] \label{Q0}\\
&-[f_T(-E)-\theta(\EF+E)]\big\},
\nonumber
\end{align}
where the step functions arise when subtracting the energy at $T=0$ because $f_{T=0}(E)=\theta(\EF-E)$. After some straightforward algebra, we find
\begin{align}
\frac{Q}{A}=\beta\,\frac{(\kB T)^3}{(\hbar\vF)^2},
\label{Q1}
\end{align}
where the thermal coefficient
\begin{widetext}
\begin{align}
\beta=\frac{2}{\pi}\left[\int_0^\infty x^2dx\left(\frac{1}{\ee^{x+\mu/\kB T}+1}+\frac{1}{\ee^{x-\mu/\kB T}+1}\right)
-\frac{1}{3}\left(\frac{\EF}{\kB T}\right)^3\right]
\label{beta}
\end{align}
\end{widetext}
explicitly depends on $\mu/\kB T$, which is in turn a function of $\EF/\kB T$ [see Eq.\ (\ref{n1})], so we find that $\beta$ is only a function of $\EF/\kB T$. Numerical evaluation of Eq.\ (\ref{beta}) yields the results shown in Fig.\ \ref{Fig7}. For $\EF\ll\kB T$, we have $\beta=(4/\pi)\int_0^\infty\theta^2d\theta/(1+\ee^\theta)\approx2.2958$. (Incidentally, we correct this parameter here for a factor of 2 that was missing in Ref.\ \cite{paper235}.) We note that the graphene heat capacity has been widely used in previous studies\cite{JKW16,SRL12,STZ14} in the so-called degenerate limit ($\kB T\ll\mu$).

\begin{widetext}

\section{Graphene conductivity}

We adopt the local-RPA model for the graphene conductivity \cite{paper235,GSC06,GSC09}
\begin{align}
\sigma(\omega)= & \frac{e^2}{\pi\hbar^2}\frac{\ii}{(\omega+\ii\tau^{-1})}\left\{\mu^{\rm D}
-\int_0^\infty dE\,\frac{f_T(E)-f_T(-E)}{1-4E^2/\left[\hbar^2(\omega+\ii\tau^{-1})^2\right]}\right\},
\label{localRPA}
\end{align}
where
\begin{align}
\mu^{\rm D}=\mu+2\kB T \log\left(1+e^{-\mu/\kB T}\right)
\label{muD}
\end{align}
is a temperature-dependent effective Drude weight that accounts for intraband transitions and has been the object of a recent theoretical and experimental study \cite{NWG16}. The integral term in Eq.\ (\ref{localRPA}) represents the contribution from interband transitions. Besides the explicit dependence on temperature $T$, we note that there is an additional dependence through the chemical potential $\mu$. We plot the resulting $\mu^{\rm D}$ in Fig.\ \ref{Fig6} (red-solid curve). A reasonable approximation to this parameter is obtained by substituting $\EF$ for $\mu$ in Eq.\ (\ref{muD}) (dashed-blue curve in Fig.\ \ref{Fig6}).

We assume a rather conservative value for the energy broadening $\hbar\tau^{-1}=10\,$meV throughout this work (this corresponds to a Drude-model mobility \cite{AM1976} $ev_{\rm F}^2\tau/\EF=3300\,$cm$^2$V$^{-1}$s$^{-1}$ for $\EF=0.2\,$eV). For simplicity, we further neglect the dependence of $\tau$ on temperature and chemical potential, which could be readily incorporated following previous studies\cite{FLS14,JKW16,STZ14}. This dependence is partially absorbed in the assumed value of $\tau$ over the significant range of temperatures under consideration, although a more detailed analysis could reveal unexpected effects outside that range.

\section{Time evolution}

The temporal evolution of the electronic temperature is given by Eqs.\ (\ref{eq:transfer}), which we solve numerically by using a $4^{\rm th}$ order Runge-Kutta method. It is instructive to rewrite them with the temperatures appearing in a more explicit form. Using the $Q_\ell$ dependence on $T_\ell$ given by Eq.\ (\ref{Q1}), we find
\begin{align}
C(T_\ell)\,\dot{T}_\ell=-\frac{T_\ell}{\tau_{\rm ph}}+\frac{4\hbar^2v_{\rm F}^2}{\pi D_\ell^2k_{\rm B}^3T_\ell^2}P_\ell(T_1,T_2),
\nonumber
\end{align}
where $C(T_\ell)=3+(T_\ell/\beta)(d\beta/dT_\ell)$ is a dimensionless coefficient that varies between 3 and 4 in the large and small $T_\ell$ limits, respectively (see $\beta$ dependence on $T_\ell$ in Fig.\ \ref{Fig7}).

In the simulations of Figs.\ \ref{Fig2}(d), \ref{Fig3}, \ref{FigS11}, and \ref{FigS12} we fix the initial temperatures $T_\ell$ to prescribed values. However, in the calculation of Fig.\ \ref{Fig4} the initial temperatures are determined by the energy absorbed from a light pulse via the absorption cross-section given by Eq.\ (\ref{sigmaabs}). Assuming a $\delta$-function pulse of frequency $\omega_0$ and fluence $F_0$, we have $Q_\ell(t=0)=\sigma_\ell^{\rm abs}(\omega_0)F_0$. The initial temperature is then obtained by entering this value of $Q_\ell$ in Eq.\ (\ref{Q1}).

\begin{figure*}
\begin{centering}
\includegraphics[width=0.7\textwidth]{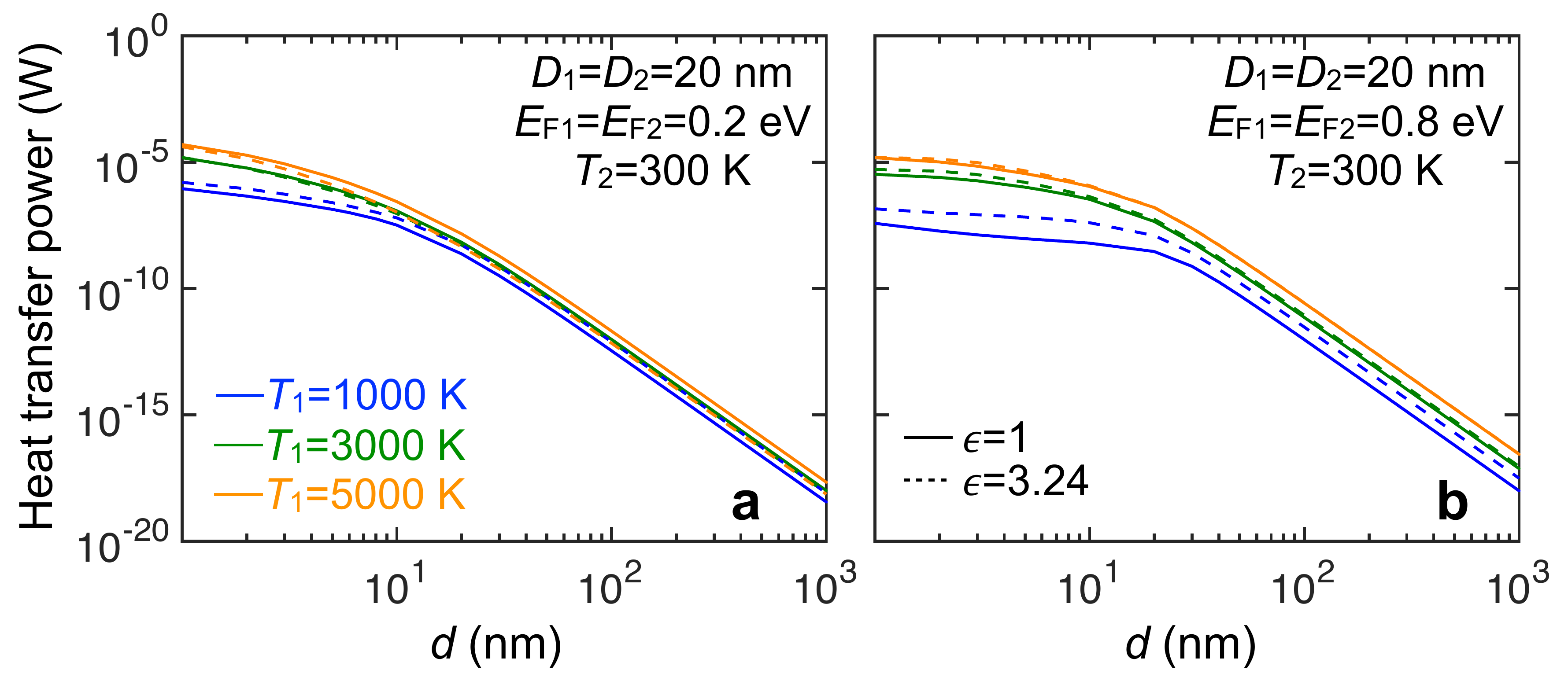}
\par\end{centering}
\caption{{\bf Influence of dielectric environment on the radiative heat transfer power (HTP).} We show the dependence of the HTP on the separation distance $d$ between two graphene nanodisks (20\,nm diameter) for different values of $T_1$ (see legend) and fixed $T_2=300\,$K. The disks are doped to a Fermi energy $\EF=0.2\,$eV in (a) and 0.8\,eV in (b). We consider homogeneous media of permittivity $\epsilon=1$ (vacuum, solid curves) or $\epsilon=3.24$ (broken curves) at the thermal wavelengths under consideration. The graphene is described using the local-RPA conductivity. A damping energy $\hbar\tau^{-1}=10\,$meV is assumed in all figures, unless otherwise stated.}
\label{FigS1}
\end{figure*}

\begin{figure*}
\begin{centering}
\includegraphics[width=0.5\textwidth]{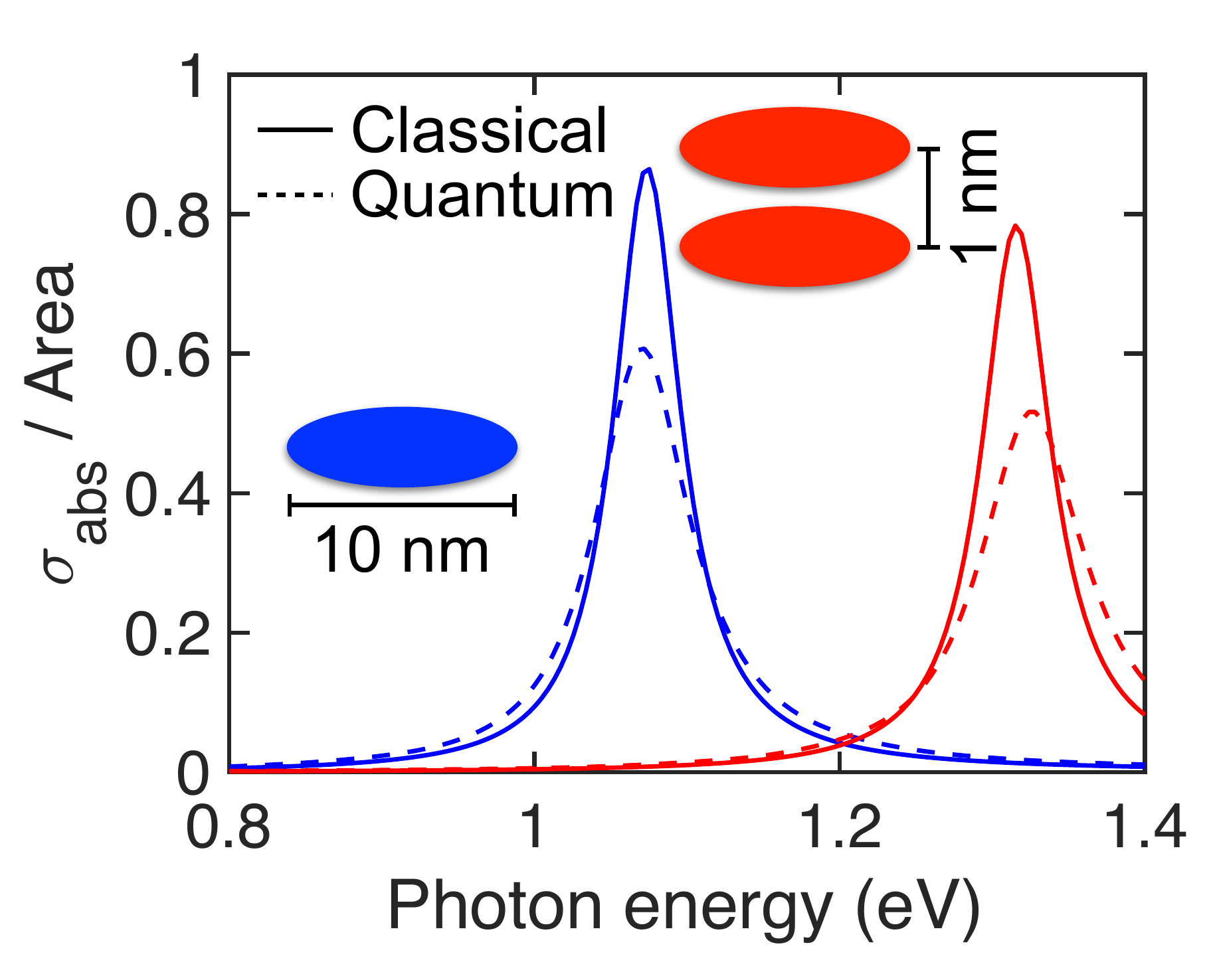}
\par\end{centering}
\caption{{\bf Influence of nonlocal effects in the response of graphene: absorption cross-section.} Absorption cross-section of individual (blue) and closely spaced (red) graphene disks calculated using either classical (local-RPA conductivity, solid curves) or quantum-mechanical (tight-binding combined with full RPA, as described elsewhere\cite{paper183}, broken curves) models. We assume a Fermi energy of 2\,eV and a damping of 0.05\,eV.}
\label{FigS2}
\end{figure*}

\begin{figure*}
\begin{centering}
\includegraphics[width=0.8\textwidth]{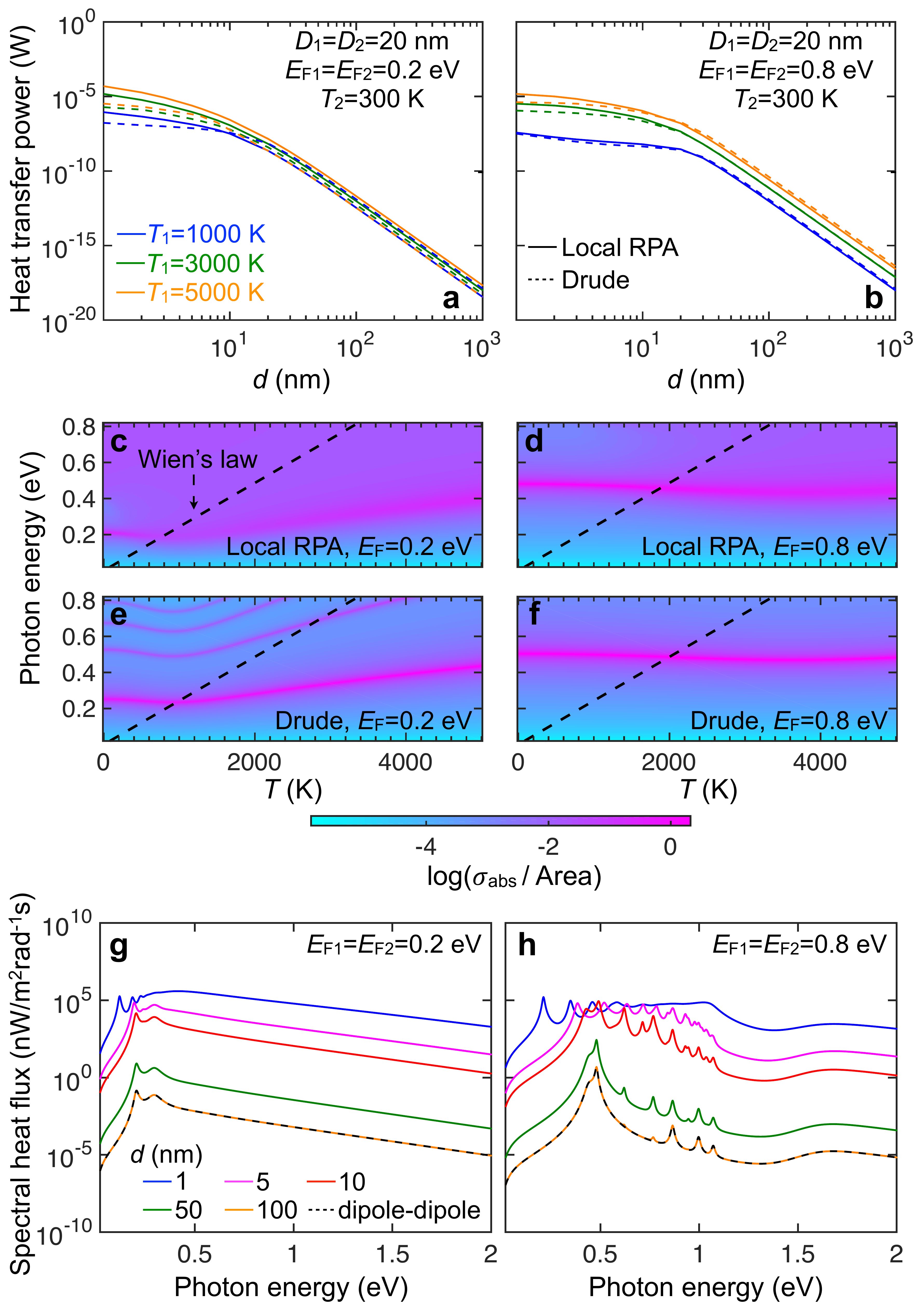}
\par\end{centering}
\caption{{\bf Dependence of the HTP on the model used for the graphene conductivity.} {\bf (a)} Dependence of the HTP on the separation distance $d$ between two graphene nanodisks (20\,nm diameter, 0.2\,eV Fermi energy) plotted for different values of $T_1$ (see legend) and fixed $T_2=300\,$K. Solid (dashed) curves are obtained with the local-RPA (Drude) model for the graphene conductivity. The temperature-dependent local-RPA is given by Eq.\ (24), while the Drude model is obtained by neglecting the integral term in that equation. {\bf (b)} Same as (a) for 0.8\,eV Fermi energy. {\bf (c-f)} Absorption cross-section of a 20\,nm graphene disk as a function of photon energy $\hbar\omega$ and temperature $T$ calculated for different values of the Fermi energy using the two models considered for the conductivity (see labels). The temperature dependence enters through the conductivity [see Eq.\ (24)]. The dashed lines correspond to Wien's law, $\hbar\omega\approx2.82\,\kB T$. {\bf (g,h)} Spectral dependence of the HTP for 20\,nm graphene disks. The vertical axis shows the value of the integrand in Eq.\ (8). The hot (cold) disk is at temperature $T_1=3000\,$K ($T_2=300\,$K). We consider different values of the disk separation $d$ and Fermi energies (see labels). The dipole-dipole approximation is shown for $d=100\,$nm (dashed curves).}
\label{FigS3}
\end{figure*}

\begin{figure*}
\begin{centering}
\includegraphics[width=0.7\textwidth]{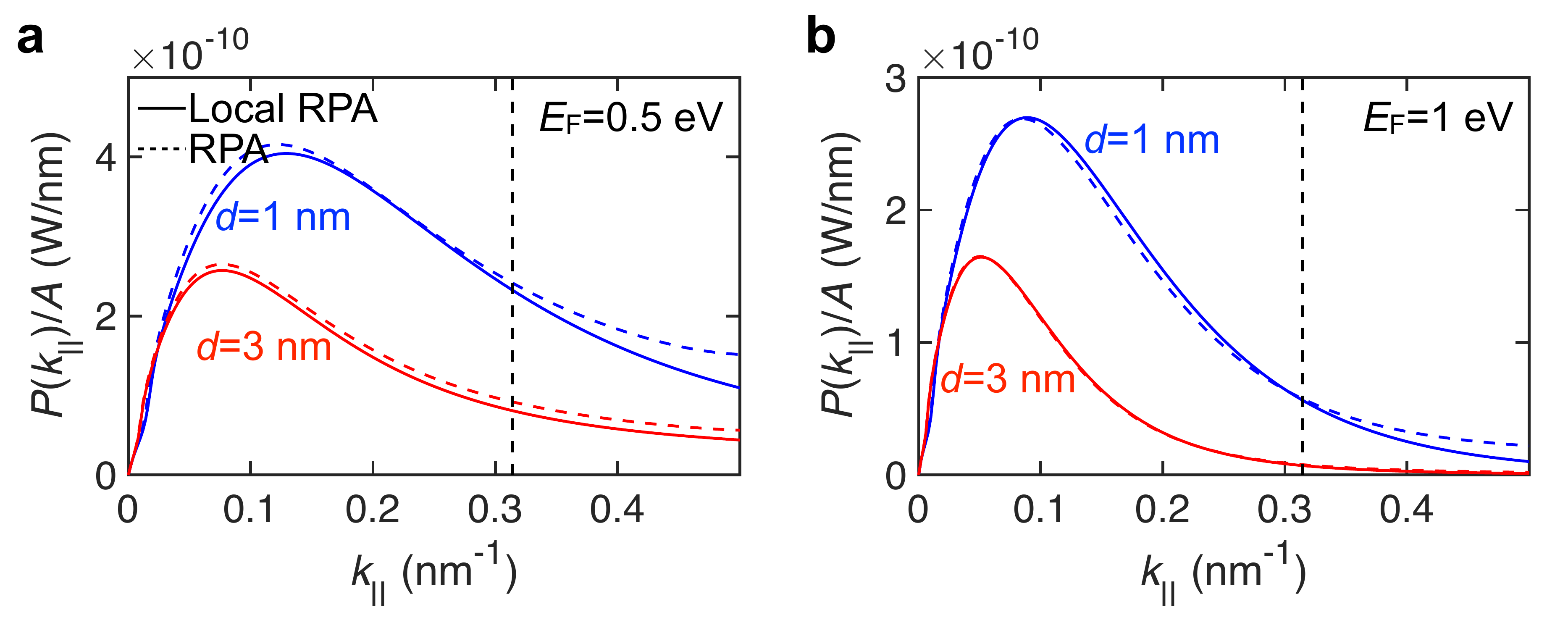}
\par\end{centering}
\caption{{\bf Influence of nonlocal effects in the response of graphene: radiative heat transfer.} We show the heat transfer power per unit area $P(k_\parallel)/A$ between two closely spaced extended graphene films resolved in parallel wave-vector components for combinations of two different Fermi energies and separations (see labels). We assume temperatures $T_1=1000\,$K and $T_2=300\,$K in the layers. The conductivity of graphene is described in the full RPA (broken curves) and in the local-RPA limit (solid curves) for a damping of 0.01\,eV. The temperature-depencence of the conductivity is neglected for simplicity. The vertical dashed lines at $k_\parallel=2\pi/D$ qualitatively indicate the region where a maximum contribution is expected for disks of diameter $D=20\,$nm.}
\label{FigS4}
\end{figure*}

\begin{figure*}
\begin{centering}
\includegraphics[width=0.7\textwidth]{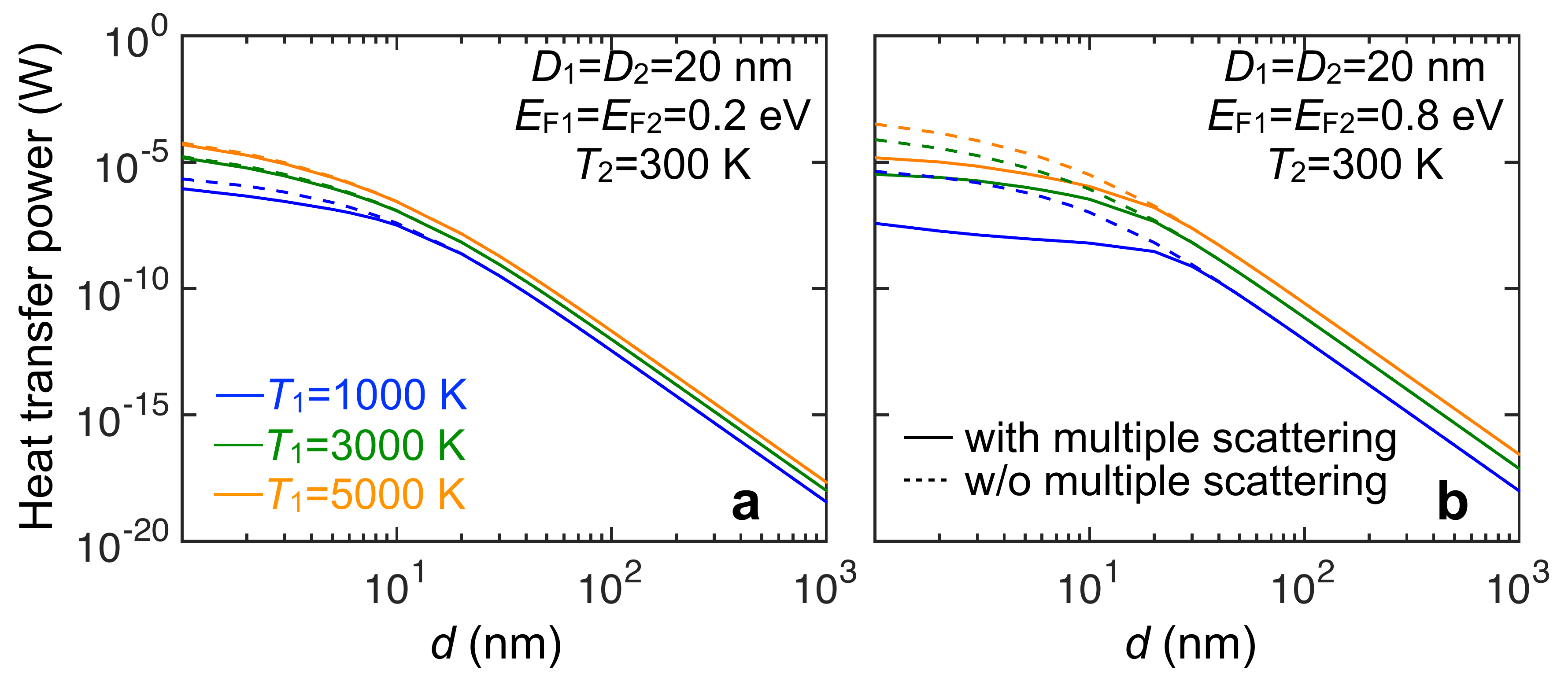}
\par\end{centering}
\caption{{\bf Influence of multiple scattering in the optical interaction between graphene disks on the HTP.} {\bf (a)} HTP under the same conditions as in Fig.\ \ref{FigS3}(a) (local-RPA) calculated with [solid curves, Eq.\ (1)] and without [dashed curves, same equation with $\Delta^m=\mathbbm{I}$] inclusion of multiple scattering in the interaction between the disks. {\bf (b)} Same as (a) for 0.8\,eV Fermi energy.}
\label{FigS5}
\end{figure*}

\begin{figure*}
\begin{centering}
\includegraphics[width=0.7\textwidth]{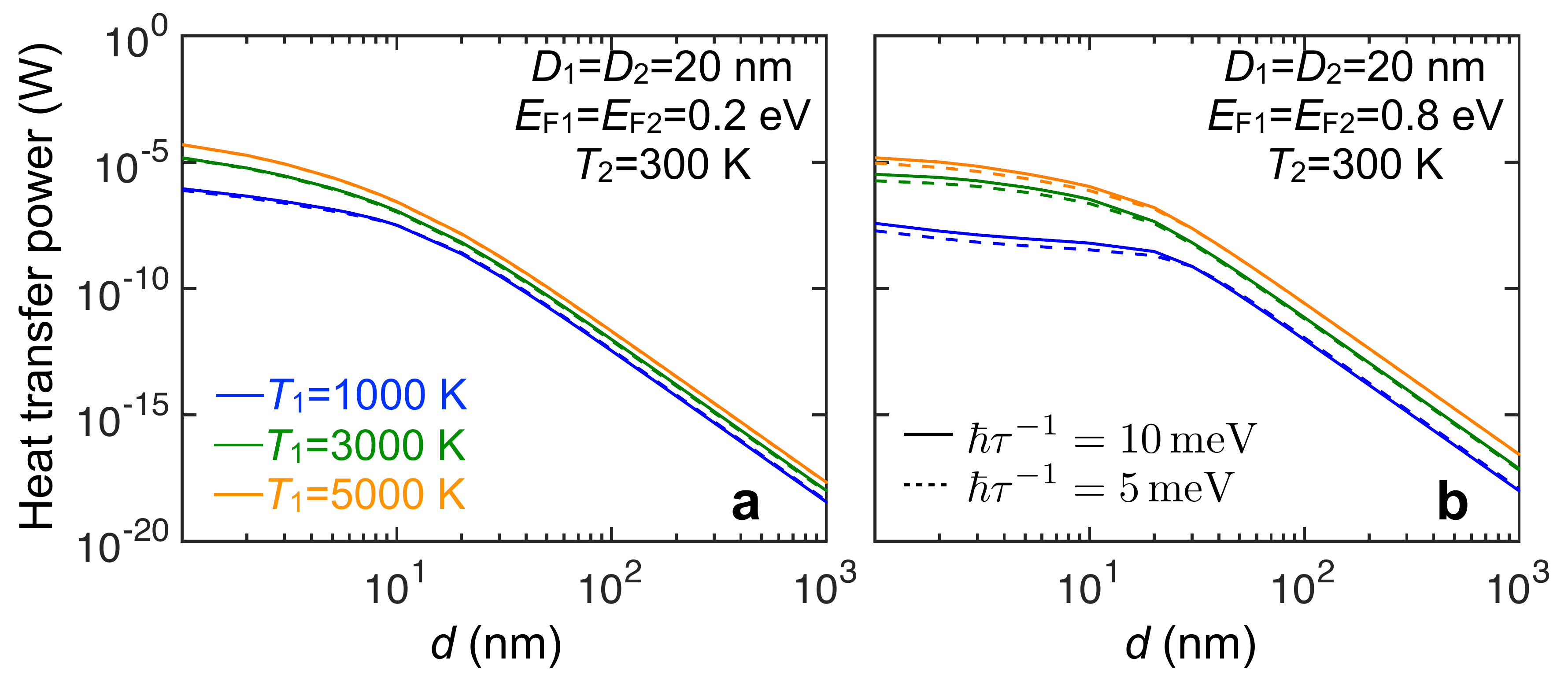}
\par\end{centering}
\caption{{\bf Influence of electronic damping on the HTP.} We show the HTP under the same conditions as in Fig.\ \ref{FigS3}(a) (local-RPA) for two different values of the inelastic broadening $\hbar\tau^{-1}$ (see labels). All other plots in this work are obtained with $\hbar\tau^{-1}=10\,$meV.}
\label{FigS6}
\end{figure*}

\begin{figure*}
\begin{centering}
\includegraphics[width=0.7\textwidth]{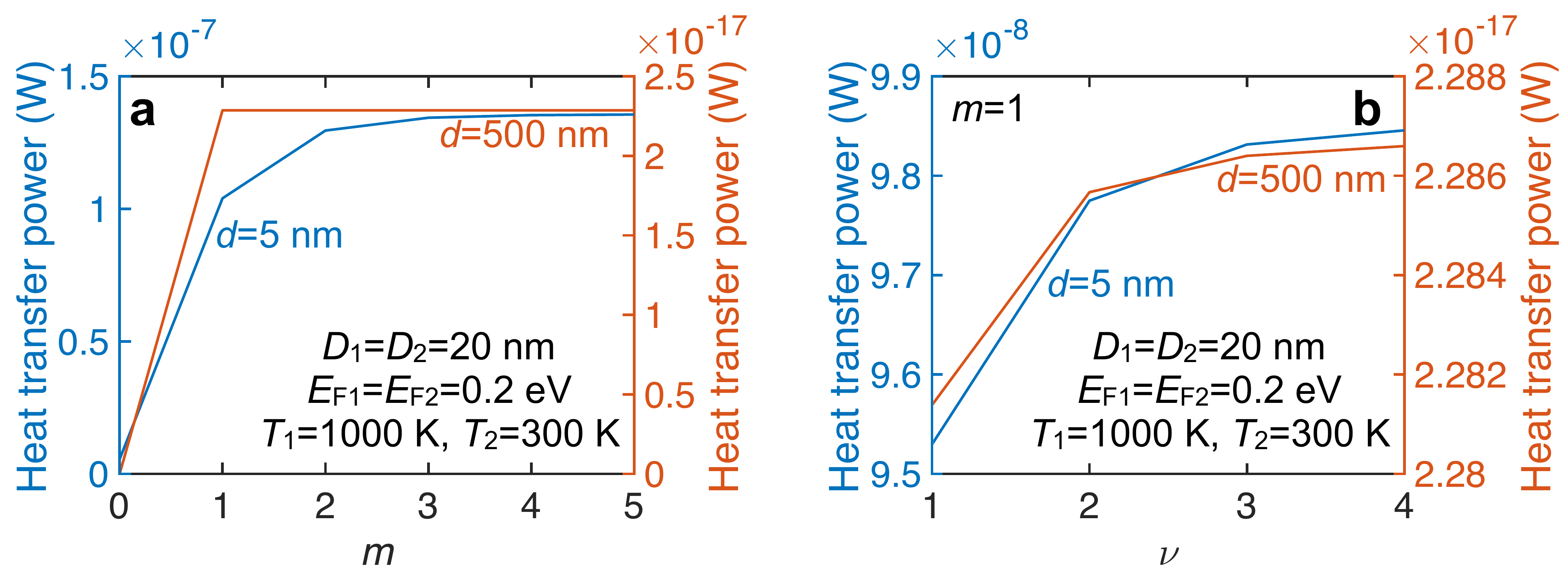}
\par\end{centering}
\caption{{\bf Convergence of the HTP with the number of PWFs.} {\bf (a)} Convergence of the HTP with increasing number of $m$'s included in the calculation under the same conditions as in Fig.\ \ref{FigS3}(a) (local-RPA) for two different disk separations $d$ (see labels). {\bf (b)} Convergence of the $m=1$ contribution to (a) as a function of the number of PWFs included in the calculation.}
\label{FigS7}
\end{figure*}

\begin{figure*}
\begin{centering}
\includegraphics[width=0.7\textwidth]{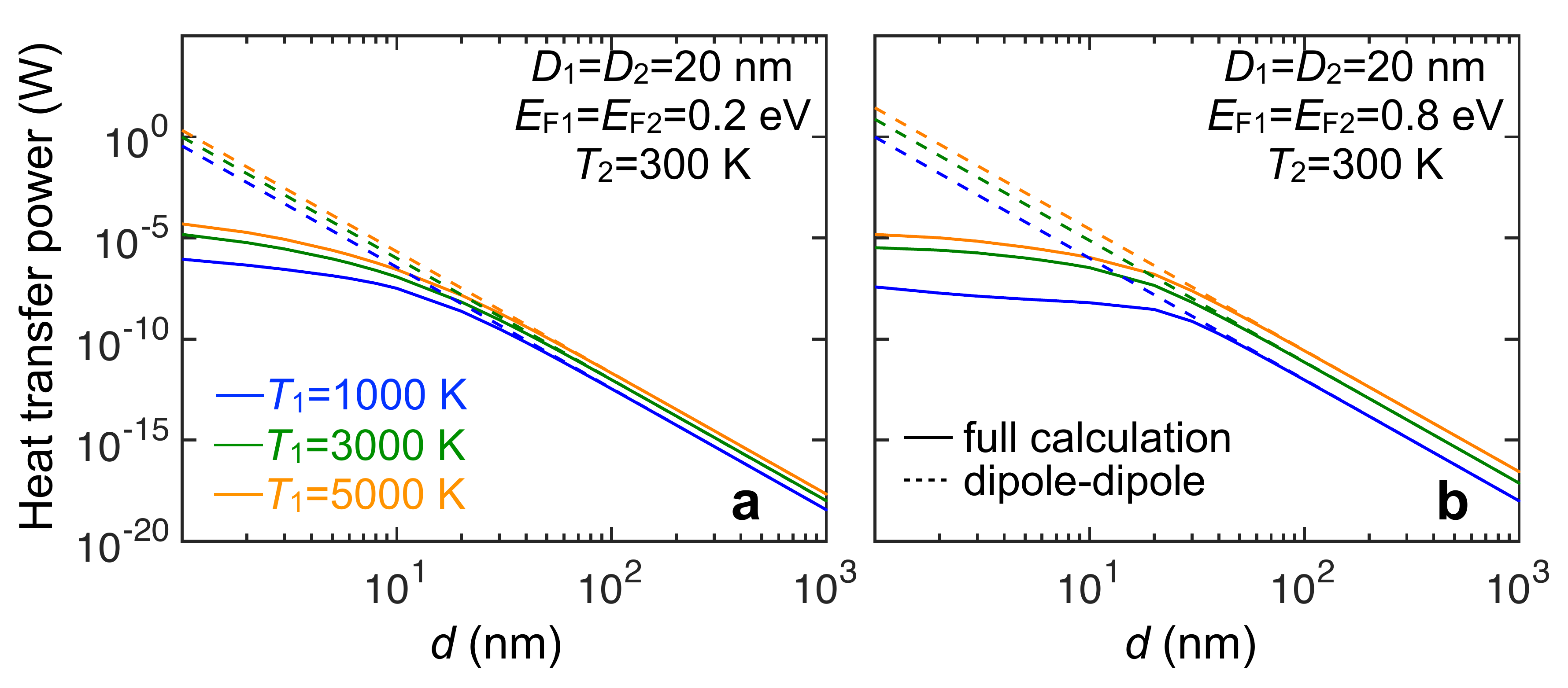}
\par\end{centering}
\caption{{\bf Influence of multipolar orders on the HTP.} We show the HTP under the same conditions as in Fig.\ \ref{FigS3}(a) (local-RPA) calculated with the full formalism [solid curves, Eq.\ (1)] and in the dipole-dipole approximation [dashed curves, Eq.\ (9)].}
\label{FigS8}
\end{figure*}

\begin{figure*}
\begin{centering}
\includegraphics[width=0.7\textwidth]{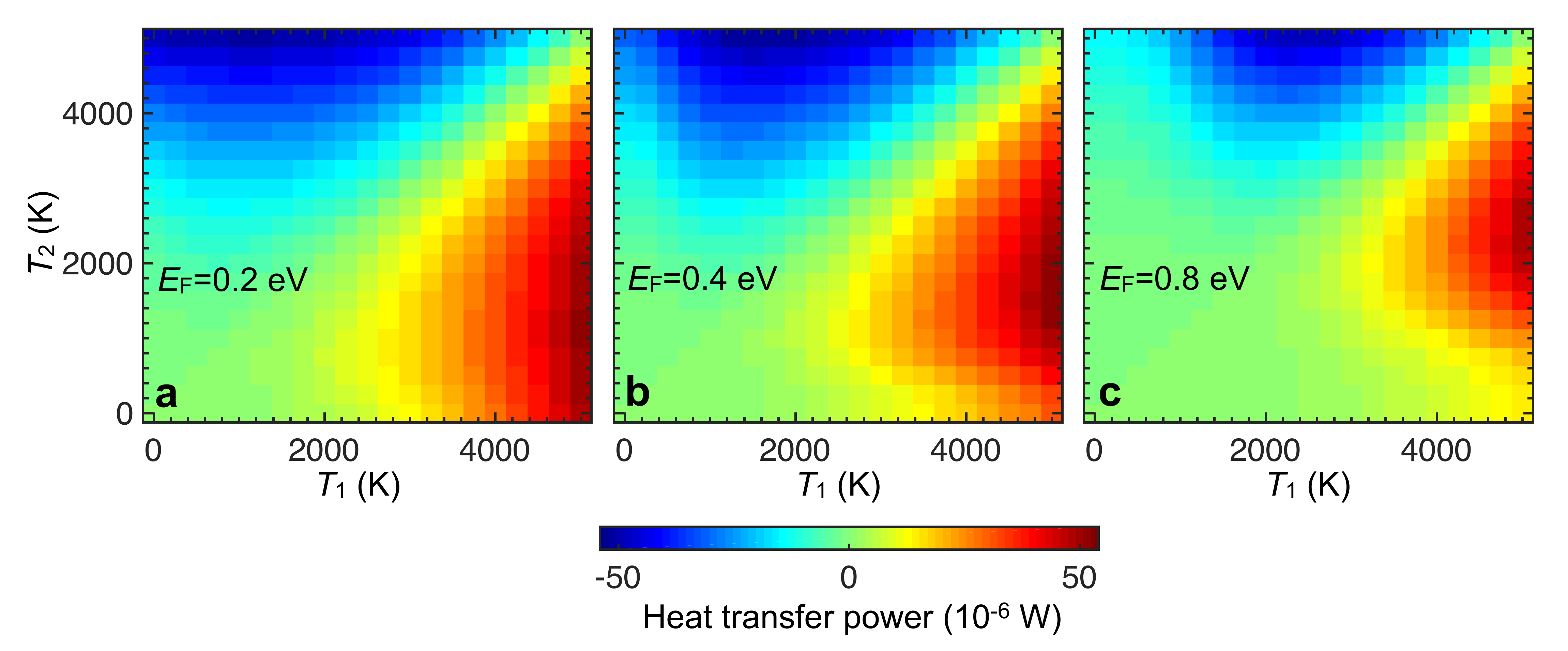}
\par\end{centering}
\caption{{\bf Temperature dependence of the HTP for different doping levels.} We consider two graphene disks of 20\,nm diameter separated by a distance $d=1\,$nm. Both disks are doped with the same Fermi energy (see labels).}
\label{FigS9}
\end{figure*}

\begin{figure*}
\begin{centering}
\includegraphics[width=0.7\textwidth]{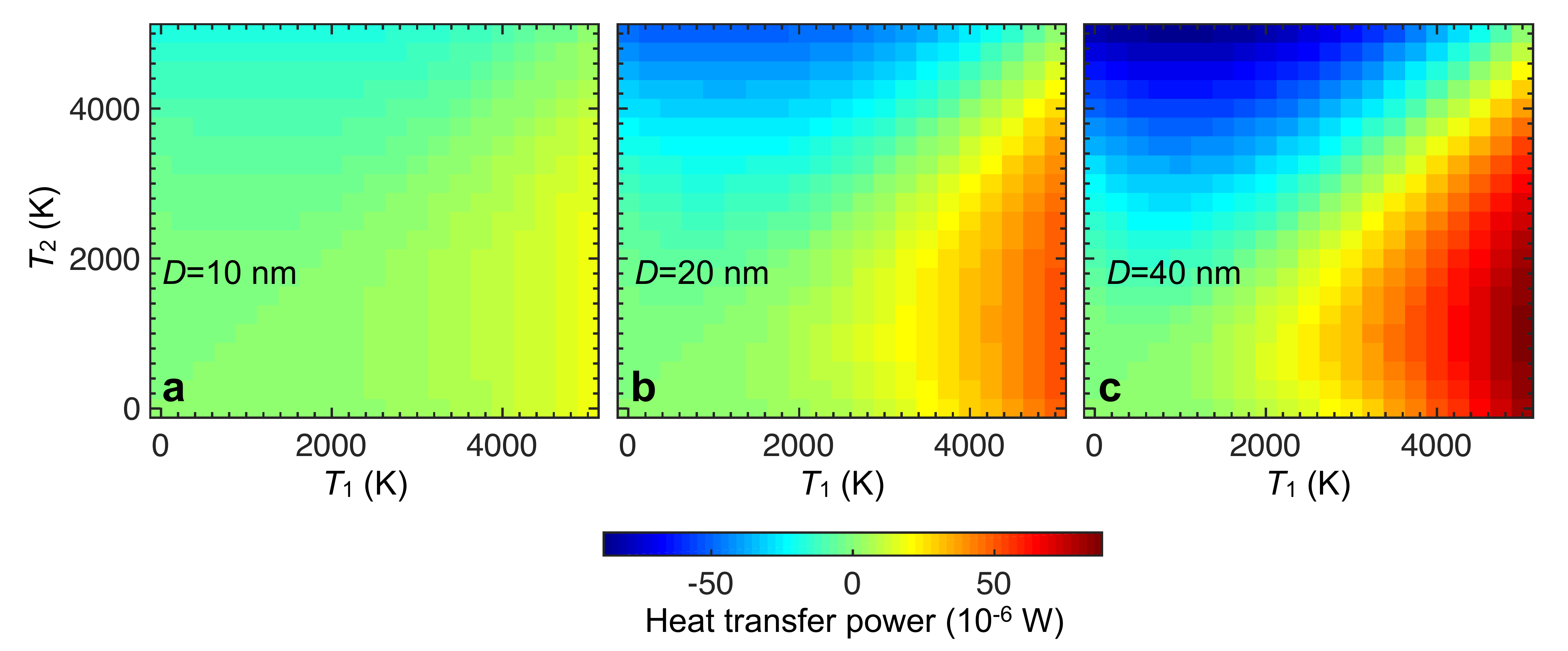}
\par\end{centering}
\caption{{\bf Temperature dependence of the HTP for different disk sizes.} Same as Fig.\ \ref{FigS9} for fixed Fermi energy $E_{{\rm F}1}=E_{{\rm F}2}=0.2\,$eV and identical disks of different size (see labels).}
\label{FigS10}
\end{figure*}

\begin{figure*}
\begin{centering}
\includegraphics[width=0.8\textwidth]{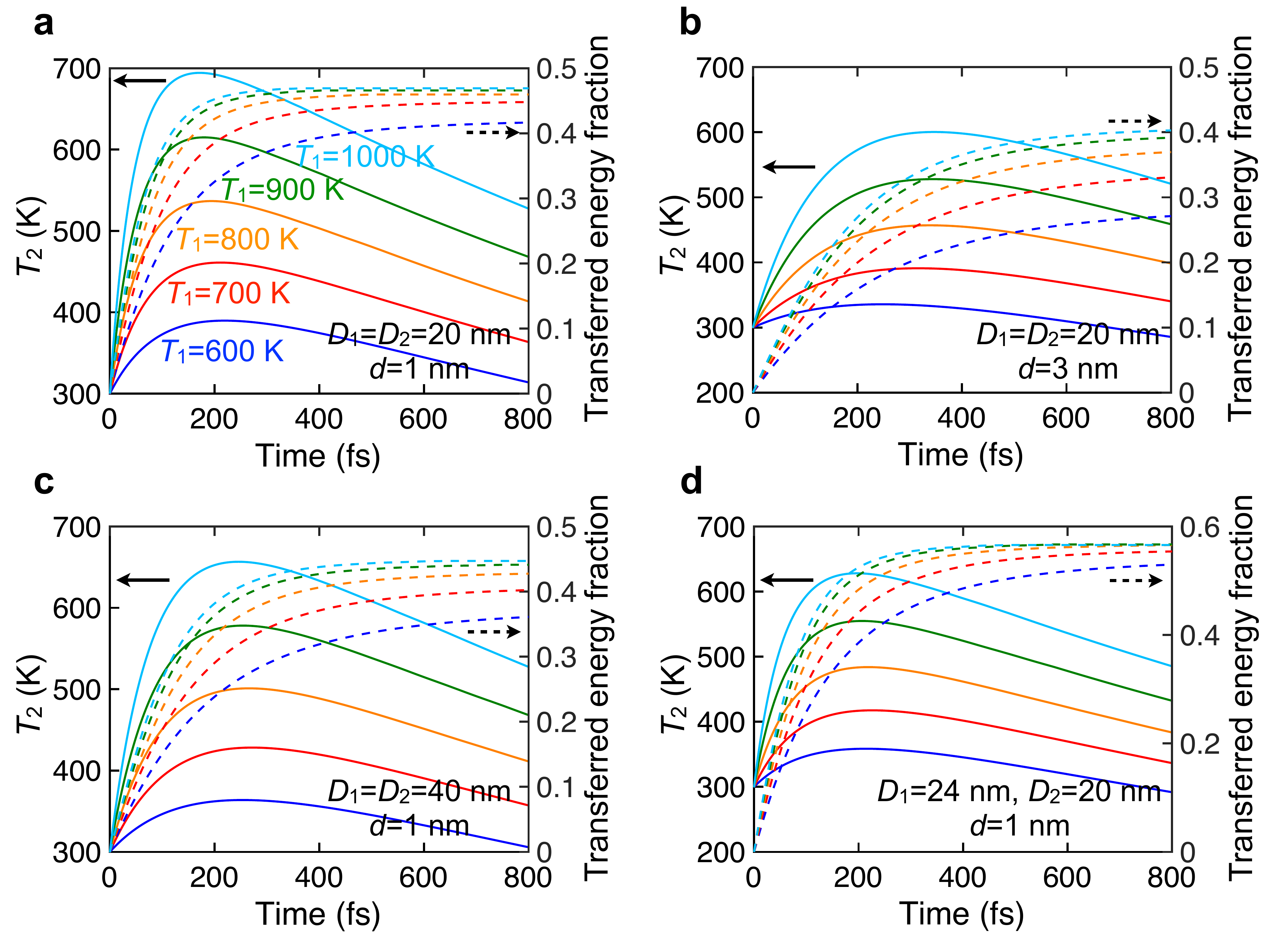}
\par\end{centering}
\caption{{\bf Size and separation dependence of the heat transfer dynamics. I.} Same as Fig.\ 3(b) for different combinations of the disk diameters $D_\ell$ and the separation $d$: 
{\bf (a)} $D_1=D_2=20\,$nm, $d=1\,$nm [same as Fig.\ 3(b)];
{\bf (b)} $D_1=D_2=20\,$nm, $d=3\,$nm (larger spacing);
{\bf (c)} $D_1=D_2=40\,$nm, $d=1\,$nm (larger disks);
{\bf (d)} $D_1=24\,$nm, $D_2=20\,$nm, $d=1\,$nm (dissimilar disks). The doping level is $\EF=0.2\,$eV in all cases.}
\label{FigS11}
\end{figure*}

\begin{figure*}
\begin{centering}
\includegraphics[width=0.8\textwidth]{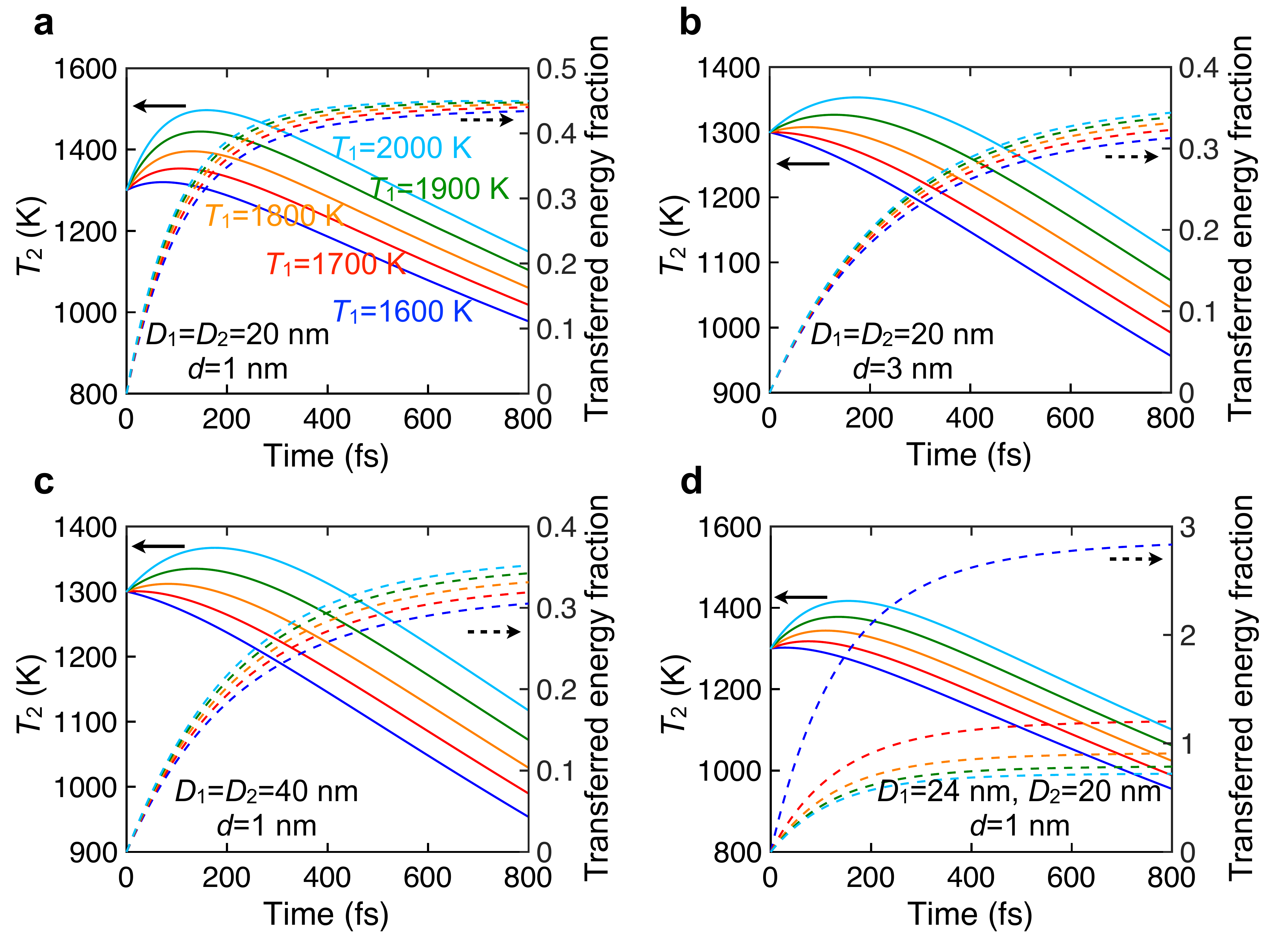}
\par\end{centering}
\caption{{\bf Size and separation dependence of the heat transfer dynamics. II.} Same as Fig.\ \ref{FigS11} for $\EF=0.5\,$eV and different initial temperatures.}
\label{FigS12}
\end{figure*}

\begin{figure*}
\begin{centering}
\includegraphics[width=0.7\textwidth]{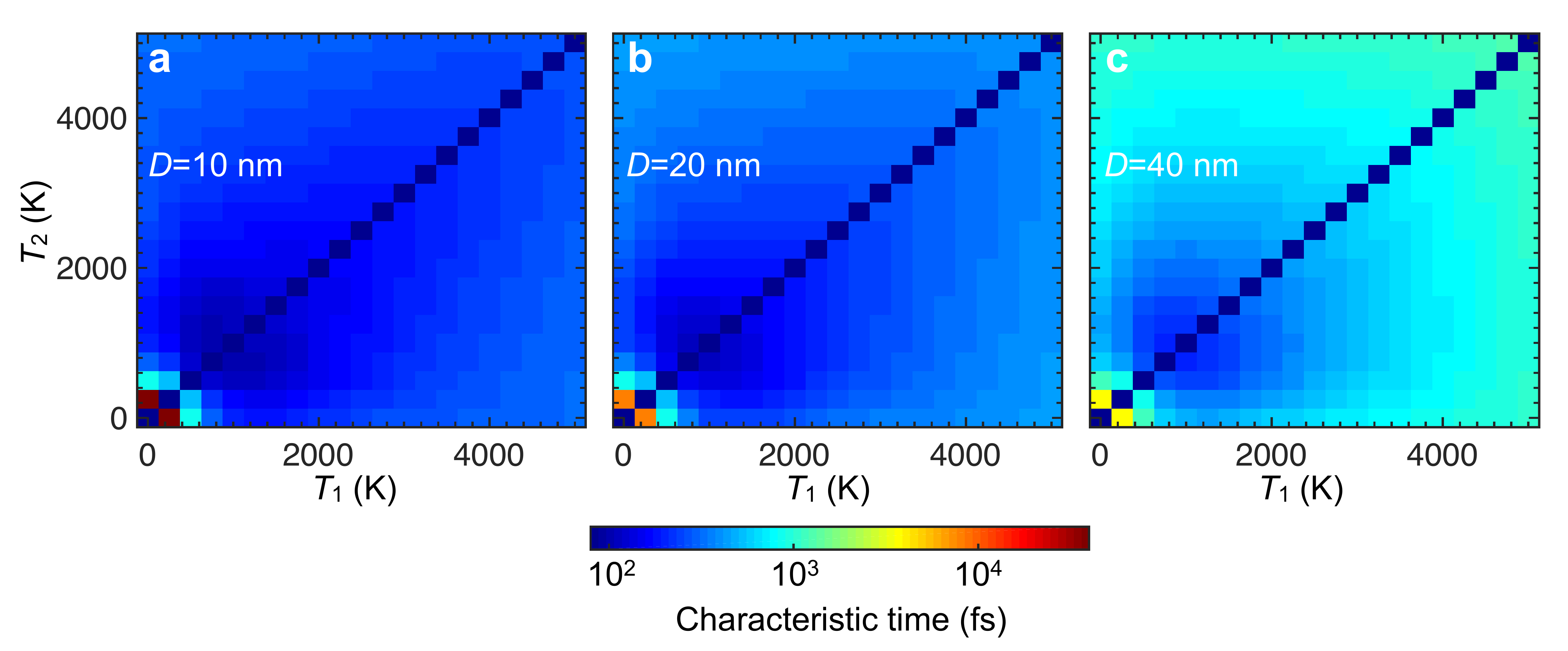}
\par\end{centering}
\caption{{\bf Temperature and size dependence of the characteristic time associated with radiative heat transfer.} We plot the heat transfer time under the same conditions as in Fig.\ \ref{FigS10}.}
\label{FigS13}
\end{figure*}

\section{Derivation of Eq.\ (6)}
\label{SN1}

We provide a brief derivation of the fluctuation-dissipation theorem \cite{N1928,CW1951} (FDT)  for fluctuations of the charge density in the frequency domain $\rho^{\rm fl}(\rb,\omega)$. We start by considering a system characterized by its charge density $\rho(\rb,t)$ and described through the Hamiltonian $H=H_0+H_1$, where $H_0$ is the unperturbed term, while
\begin{equation}
H_1=\int d^3\rb\, \rho(\rb,t)\phi(\rb,t)
\end{equation}
accounts for the time-dependent interaction between $\rho$ and an external electric potential $\phi(\rb,t)$. Using first-order perturbation theory under the assumption that $H_1$ vanishes in the $t\rightarrow-\infty$ limit, we can write the eigenstates of the perturbed system as
\begin{equation}
\ket{\psi_m(t)}\approx\ket{m}-\frac{i}{\hbar}\int_{-\infty}^t dt'\,H_1(t')\ket{m},
\end{equation}
where $\ket{m}$ is the eigenstate of $H_0$ with energy $E_m$ (i.e., $H_0\ket{m}=E_m\ket{m}$). Summing the contributions from all perturbed states $\ket{\psi_m(t)}$, we obtain the expectation value of the charge density induced by $H_1$ as
\begin{align}
\braket{\rho^{\rm ind}(\rb,t)}=&\braket{ \rho(\rb,t)}-\braket{ \rho(\rb,-\infty)} \nonumber\\
=& -\frac{i}{\hbar}\int^t_{-\infty}dt'\int d^3\rb \sum_m\frac{\ee^{-E_m/k_BT}}{Z} \bra{m}\left[\rho(\rb,t),\rho(\rb',t')\right]\ket{m}\phi(\rb',t') \nonumber\\
=& -\frac{i}{\hbar}\int^t_{-\infty}dt'\int d^3\rb\; \chi(\rb, \rb', t') \phi(\rb',t'),
\end{align}
where $Z=\sum_m\ee^{-E_m/k_BT}$ is the partition function at temperature $T$, while $ \chi(\rb, \rb', t')$ is the electric susceptibility of the system. The latter can be expressed in the frequency domain by taking the Fourier transform of the above expressions:
\begin{equation}\label{chiFDT}
\chi(\rb, \rb', \omega)=\int dt\; \chi(\rb, \rb', t)\ee^{i\omega t} = \frac{1}{Z} \sum_{m,n} \bra{m}\rho(\rb)\ket{n}  \bra{n}\rho(\rb')\ket{m}\frac{\ee^{-E_m/k_BT}-\ee^{-E_n/k_BT}}{\hbar\omega+E_m-E_n+i0^+}, 
\end{equation}
where we have used $\rho(\rb,t)= \ee^{i H_0t/\hbar}\rho(\rb)\ee^{-i H_0t/\hbar}$ as well as the closure relation $\sum_n\ket{n}\bra{n}=\mathbbm{I}$. 

At this point, we follow a similar procedure for calculating the self correlations of the fluctuating charge density $\rho^{\rm fl}(\rb,\omega)$. We find
\begin{align}
\braket{\rho^{\rm fl}(\rb,\omega)\rho^{\rm fl}(\rb',\omega')}&=\int dt dt'\; \ee^{i\omega t} \ee^{i\omega' t'} \braket{\rho^{\rm fl}(\rb,t)\rho^{\rm fl}(\rb',t')} \nonumber \\
&=\frac1{Z}\int dt dt'\; \ee^{i\omega t} \ee^{i\omega' t'}\sum_{m,n}\ee^{-E_m/k_B T}\ee^{i(E_m-E_n)(t-t')/\hbar} \bra{m}\rho(\rb)\ket{n}  \bra{n}\rho(\rb')\ket{m} \nonumber \\ &= 2\pi\delta(\omega+\omega')S(\omega),
\end{align}
where
\begin{equation}
S(\omega)=\frac{2\pi\hbar}{Z}\sum_{m,n}\ee^{-E_m/k_B T} \bra{m}\rho(\rb)\ket{n}  \bra{n}\rho(\rb')\ket{m} \delta(\hbar\omega+E_m-E_n).
\end{equation}
Comparing this expression with Eq.\ (\ref{chiFDT}), we obtain $S(\omega)=-2\hbar\;[n(\omega)+1]\;{\rm Im}\{\chi(\rb, \rb', \omega)\}$, where $n(\omega)=[\ee^{\hbar\omega/k_BT}-1]^{-1}$ is the Bose-Einstein distribution function. We conclude that
\begin{equation}
\braket{\rho^{\rm fl}(\rb,\omega)\rho^{\rm fl}(\rb',\omega')}=-4\pi\hbar \delta(\omega+\omega')\;[n(\omega)+1]\;{\rm Im}\{\chi(\rb',\rb,\omega)\}.
\end{equation}
Additionally, interchanging $\rho^{\rm fl}(\rb,\omega)$ and $\rho^{\rm fl}(\rb',\omega')$, we have
\begin{equation}
\braket{\rho^{\rm fl}(\rb,\omega)\rho^{\rm fl}(\rb',\omega')}=-4\pi\hbar \delta(\omega+\omega')\;n(\omega)\;{\rm Im}\{\chi(\rb',\rb,\omega)\}.
\end{equation}
Finally, noting that $\chi(\rb,\rb',\omega)=\chi(\rb',\rb,\omega)$, the expectation value of the physically meaningful symmetrized correlation becomes
\begin{equation}
\braket{\rho^{\rm fl}(\rb',\omega')\rho^{\rm fl}(\rb,\omega)}_{\rm sym}=\frac1{2}\left[\braket{\rho^{\rm fl}(\rb,\omega)\rho^{\rm fl}(\rb',\omega')}+\braket{\rho^{\rm fl}(\rb',\omega')\rho^{\rm fl}(\rb,\omega)}\right]=-4\pi\hbar \delta(\omega+\omega')\; [n(\omega)+\frac1{2}]\;{\rm Im}\{\chi(\rb',\rb,\omega)\}.
\end{equation}
This is the FDT used in Eq.\ (6), where we drop the 'sym' subscript for clarity.

\section{Derivation of Eqs.\ (7) and (8)}
\label{SN2}

We start from Eq.\ (4), which we recast as
\begin{align}
P_{1\leftarrow2}=\ii&\iint\frac{d\omega d\omega'}{(2\pi)^2}
\,\omega\,
\ee^{-\ii(\omega+\omega')t}
\int d^3\rb\int d^3\rb'
\left\langle
\left[\left(\rho_2^{\rm fl}(\omega)\right)^{\rm T}\cdot\Delta^{\rm T}(\omega)\cdot v\cdot\chi_1(\omega)\cdot v\cdot
\Delta(\omega')
\cdot\rho_2^{\rm fl}(\omega')\right]\bigg|_{\rb,\rb'}
\right\rangle.
\end{align}
This expression can be conveniently rewritten by moving $\left(\rho_2^{\rm fl}(\omega)\right)^{\rm T}$ to the right end as
\begin{align}
P_{1\leftarrow2}=\ii&\iint\frac{d\omega d\omega'}{(2\pi)^2}
\,\omega\,
\ee^{-\ii(\omega+\omega')t}
\int d^3\rb\int d^3\rb'
\left[\Delta^{\rm T}(\omega)\cdot v\cdot\chi_1(\omega)\cdot v\cdot
\Delta(\omega')
\cdot\left\langle\rho_2^{\rm fl}(\omega')\cdot\left(\rho_2^{\rm fl}(\omega)\right)^{\rm T}\right\rangle\right]\bigg|_{\rb,\rb'}.
\end{align}
Here, $\rho_2^{\rm fl}(\omega')\cdot\left(\rho_2^{\rm fl}(\omega)\right)^{\rm T}$ is a matrix formed by the product of column and row vectors. Charge fluctuations are readily evaluated using the FTD [Eq.\ (6)] together with the identity $\chi_\ell(\omega)=\chi_\ell^*(-\omega)$. We find
\begin{align}
P_{1\leftarrow2}=\frac{-i\hbar}{\pi}\int\omega\,d\omega
\,(n_2+1/2) \;{\rm Tr}\left[
\Delta^{\rm T}\cdot v\cdot\chi_1\cdot v\cdot\Delta^*\cdot{\rm Im}\{\chi_2\}
\right]
\nonumber\\
=\frac{-i\hbar}{\pi}\int\omega\,d\omega
\,(n_2+1/2) \;{\rm Tr}\left[
{\rm Im}\{\chi_2\}\cdot\Delta^\dagger\cdot v\cdot\chi_1\cdot v\cdot\Delta
\right],
\end{align}
where the second line is obtained from the first one by applying the matrix trace identity ${\rm Tr}[A]={\rm Tr}[A^{\rm T}]$ as well as $v=v^{\rm T}$ and $\chi_\ell=\chi_\ell^{\rm T}$. We note that a dependence of $\chi_\ell$, $\Delta$, and $n_2$ on $\omega$ is understood. We now split the integral as $\int d\omega\rightarrow\int_0^\infty d\omega+\int_{-\infty}^0d\omega$ and change $\omega$ to $-\omega$ in the negative frequency term. Using the property $[n_\ell(\omega)+1/2]=-[n_\ell(-\omega)+1/2]$, we obtain
\begin{align}
P_{1\leftarrow2}=\frac{-i\hbar}{\pi}\int_0^\infty\omega\,d\omega
\,(n_2+1/2) \;{\rm Tr}\left[
{\rm Im}\{\chi_2\}\cdot\Delta^\dagger\cdot v\cdot\chi_1\cdot v\cdot\Delta
-
{\rm Im}\{\chi_2\}\cdot\Delta^{\rm T}\cdot v\cdot\chi_1^*\cdot v\cdot\Delta^*
\right].
\label{aaa}
\end{align}
Taking the transpose of the second term and using the above matrix properties together with ${\rm Tr}[A\cdot B]={\rm Tr}[B\cdot A]$, Eq.\ (\ref{aaa}) reduces to
\begin{align}
P_{1\leftarrow2}&=\frac{-i\hbar}{\pi}\int_0^\infty\omega\,d\omega
\,(n_2+1/2) \;{\rm Tr}\left[
{\rm Im}\{\chi_2\}\cdot\Delta^\dagger\cdot v\cdot\chi_1\cdot v\cdot\Delta
-
\Delta^\dagger\cdot v\cdot\chi_1^*\cdot v\cdot\Delta\cdot{\rm Im}\{\chi_2\}
\right]
\nonumber\\
&=\frac{-i\hbar}{\pi}\int_0^\infty\omega\,d\omega
\,(n_2+1/2) \;{\rm Tr}\left[
\Delta^\dagger\cdot v\cdot\chi_1\cdot v\cdot\Delta\cdot{\rm Im}\{\chi_2\}
-
\Delta^\dagger\cdot v\cdot\chi_1^*\cdot v\cdot\Delta\cdot{\rm Im}\{\chi_2\}
\right]
\nonumber\\
&=\frac{-i\hbar}{\pi}\int_0^\infty\omega\,d\omega
\,(n_2+1/2) \;{\rm Tr}\left[\Delta^\dagger\cdot v\cdot\left(\chi_1-\chi_1^*\right)\cdot v
\cdot\Delta\cdot{\rm Im}\{\chi_2\}\right]
\nonumber\\
&=\frac{2\hbar}{\pi}\int_0^\infty\omega\,d\omega
\,(n_2+1/2) \;{\rm Tr}\left[\Delta^\dagger\cdot v\cdot{\rm Im}\{\chi_1\}\cdot v
\cdot\Delta\cdot{\rm Im}\{\chi_2\}\right],
\end{align} 
which is Eq.\ (7). A similar argument can be followed to prove that $P_{1\leftarrow2}$ is indeed a real number.

When interchanging the subindices $1\Leftrightarrow2$, upon inspection of the definition of $\Delta$ [Eq.\ (5)], we have $v\cdot\Delta\Leftrightarrow\Delta^{\rm T}\cdot v$. Using this transformation, as well as the trace properties noted above, we find that the expression in the square brackets of Eq.\ (7) remains the same upon index interchange. This directly leads to Eq.\ (8) for the difference $P_{2\leftarrow1}-P_{1\leftarrow2}$.

\section{Computation of $v_{jj'}$ for coaxial disks}
\label{SN3}

In this section, we provide a computationally efficient expression to calculate the Coulomb interaction matrix elements $v_{jj'}$ [Eq.\ (14)] for coaxial disks [i.e., with the plasmon wave functions (PWFs) of Eqs.\ (16)]. We start by rewriting the Coulomb potential as\cite{J99}
\begin{align}
\frac{1}{|\rb-\rb'|}=4\pi\sum_{l=0}^\infty\sum_{m=-l}^l\frac{1}{2l+1}\frac{r_<^l}{r_>^{l+1}} Y_{lm}(\Omega_\rb) Y_{lm}^*(\Omega_{\rb'}),
\label{vexp}
\end{align}
where $Y_L$ are spherical harmonics, $r_<={\rm min}\{r,r'\}$, and $r_>={\rm max}\{r,r'\}$. Specifying Eq.\ (14) for two PWFs $\rho_{m\nu}^\kappa(\th)$ and $\rho_{m'\nu'}^{\kappa'}(\th')$ and using Eq.\ (\ref{vexp}), we can perform the azimuthal integrals of $\th$ and $\th'$ analytically by choosing the spatial origin at a point along the axis of revolution symmetry in between the two disks. Upon detailed examination, we find $v_{jj'}$ to be zero unless $m=m'$ and $\kappa=\kappa'$. Therefore, PWFs of different azimuthal symmetry do not interact. It should be also noted that only $\kappa={\rm c}$ contributes to $m=0$. The remaining nonzero elements are independent of $\kappa$, but they depend on $m$, $\nu$, and $\nu'$ as
\begin{align}
v_{\nu\nu'}^m=\left(1+\delta_{m0}\right) \;\frac{8\pi^3D_1^2D_2^2}{\epsilon} \sum_{l=m}^\infty\frac{1}{2l+1} \int_0^{1/2}\theta\,d\theta\,\rho_{m\nu}(\theta) \int_0^{1/2}\theta'd\theta'\,\rho_{m\nu'}(\theta')\;
\frac{r_<^l}{r_>^{l+1}}\;Y_{lm}(\theta_1,0)\,Y_{lm}(\theta_2,0),
\label{vmnunu}
\end{align}
where we take $r=D_1\theta$ and $r'=D_2\theta'$. Additionally, the spherical harmonics in this expression are evaluated at zero azimuthal angle, while the polar angles are $\theta_1=\tan^{-1}(D_1\theta/d_1)$ and $\theta_2=\pi-\tan^{-1}(D_2\theta'/d_2)$, where $d_1$ and $d_2$ are the distances from the disks to the origin (i.e., $d_1+d_2=d$), a convenient choice being $d_1=d$ and $d_2=0$, so that $(r_</r_>)^l$ goes rapidly down for large $l$, particularly at large separations. Equation\ (\ref{vmnunu}) gives the $(\nu\nu')$ elements of the matrix $v^m$ entering Eq.\ (1). This expression is also useful to normalize the PWFs via Eq.\ (13), whose integral corresponds to $v^m_{\nu\nu'}$ with $\epsilon=1$, $D_1=D_2$, and $d=0$.

\section{Radiative heat transfer between extended graphene films}
\label{SN4}

The lack of translational invariance in graphene disks prevents us from including nonlocal effect in the classical description of their optical response. In order to assess the relative contribution of such effects, we consider extended graphene films, for which the nonlocal conductivity admits analytical expressions\cite{WSS06,HD07}. The radiative heat transfer power can then be decomposed in components associated with different parallel wave vectors $\kb_\parallel$. We argue that the relative importance of nonlocal contributions for a disk of diameter $D$ is roughly the same as for the $k_\parallel=2\pi/D$ component in the extended films. An expression for the transfer power between films can be obtained by starting from Eq.\ (8), replacing the trace by the sum $\sum_{\kb_\parallel}\rightarrow(A/4\pi^2)\int d^2\kb_\parallel$, where $A$ is the film area, and writing $v\rightarrow2\pi/k_\parallel$ for the Coulomb interaction in $\kb_\parallel$ space. Additionally, from a direct analysis of the electrostatic problem, we have $v\cdot\chi_\ell\rightarrow-r_\ell$, where $r_\ell=1/(1-\ii\omega/2\pi k_\parallel\sigma_\ell)$ is the graphene reflection coefficient for TM polarization (notice that the reflection for TE polarization vanishes in the quasistatic limit), while $\sigma_\ell$ is the conductivity of the layer $\ell=1,2$. Putting these elements together, the transfer power per unit area becomes $\int_0^\infty dk_\parallel P(k_\parallel)/A$, where
\begin{align}
\frac{1}{A}P(k_\parallel)
=\frac{\hbar k_\parallel}{\pi^2} \int_0^\infty\omega d\omega (n_1-n_2)\ee^{-2k_\parallel d}\frac{{\rm Im}\{r_1\}{\rm Im}\{r_2\}}{\left|1-r_1r_2\exp(-2k_\parallel d)\right|^2},
\end{align}
in agreement with the $c\rightarrow\infty$ limit of the well-known expression for the transfer power between two planar structures\cite{SVC12}. We plot this quantity in Fig.\ \ref{FigS4} using the full nonlocal RPA (broken curves) and the local-RPA (solid curves) models for the conductivity. The agreement between these results indicates that nonlocal effects only play a marginal role in this study.

\end{widetext}


\section*{Acknowledgments}

We thank J. R. M. Saavedra for help in the design of Fig.\ \ref{Fig1}. This work has been supported in part by the Spanish MINECO (MAT2014-59096-P and SEV2015-0522), AGAUR (2014 SGR 1400), Fundaci\'o Privada Cellex, the European Commission (Graphene Flagship CNECT-ICT-604391 and FP7-ICT-2013-613024-GRASP), and Dept. Physics and Astronomy and College of Arts and Sciences of the University of New Mexico.


\end{document}